\documentclass[apj]{emulateapj}

\newcommand{\co}          {{\rm $^{12}$CO}}
\newcommand{\hi}          {\mbox{\rm \ion{H}{1}}}
\newcommand{\hii}         {\mbox{\rm \ion{H}{2}}}
\newcommand{\jone}        {($J=1\rightarrow0$)}
\newcommand{\kms}         {km~s$^{-1}$}
\newcommand{\gpercmcu}    {g~cm$^{-3}$}
\newcommand{\ha}          {\mbox{H$\alpha$}}
\newcommand{\mlk}         {\mbox{$M_{\odot}/L_{\odot, K}$}}
\newcommand{\mlr}         {\mbox{$M_{\odot}/L_{\odot, R}$}}
\newcommand{\mlrs}        {\mbox{$M_{\odot}/L_{\odot, r'}$}}
\newcommand{\mslk}        {\mbox{$M_{*}/L_{K}$}}
\newcommand{\msl}         {\mbox{$M_{*}/L$}}
\newcommand{\mslr}        {\mbox{$M_{*}/L_{R}$}}
\newcommand{\mslrs}       {\mbox{$M_{*}/L_{r'}$}}
\newcommand{\hr}          {\mbox{$^h$}}
\renewcommand{\min}       {\mbox{$^m$}}
\newcommand{\rotcur}      {{\sc rotcur}}
\newcommand{\ringfit}     {{\sc ringfit}}
\newcommand{\alphadm}     {\alpha_{\mbox{{\tiny DM}}}}
\newcommand{\rhodm}       {\rho_{\mbox{{\tiny DM}}}}

\shorttitle{Dark Matter Density Profiles}
\shortauthors{Simon et al.}
\slugcomment{Accepted for publication in \emph{The Astrophysical Journal}}

\begin{document}

\title{High-Resolution Measurements of the Halos of Four Dark
Matter-Dominated Galaxies: Deviations From a Universal Density
Profile\altaffilmark{1}}

\author{Joshua D. Simon\altaffilmark{2}, Alberto
D. Bolatto\altaffilmark{2}, Adam Leroy\altaffilmark{2}, and Leo Blitz}

\altaffiltext{1}{Based on observations carried out at the WIYN
Observatory.  The WIYN Observatory is a joint facility of the
University of Wisconsin-Madison, Indiana University, Yale University,
and the National Optical Astronomy Observatory.}

\altaffiltext{2}{Visiting Astronomer, Kitt Peak National Observatory,
National Optical Astronomy Observatory, which is operated by the
Association of Universities for Research in Astronomy, Inc. (AURA)
under cooperative agreement with the National Science Foundation.  }

\affil{Department of Astronomy, 601 Campbell Hall, University of 
       California at Berkeley, CA  94720}
\email{jsimon@astro.berkeley.edu}
\email{bolatto@astro.berkeley.edu}
\email{aleroy@astro.berkeley.edu}
\email{blitz@astro.berkeley.edu}

\medskip
\and
\vspace{-0.3cm}

\author{Elinor L. Gates}

\affil{Lick Observatory, P.O. Box 85, Mount Hamilton, CA  95140}
\email{egates@ucolick.org}

\begin{abstract}

We derive rotation curves for four nearby, low-mass spiral galaxies
and use them to constrain the shapes of their dark matter density
profiles.  This analysis is based on high-resolution two-dimensional
\ha\ velocity fields of NGC~4605, NGC~5949, NGC~5963, and NGC~6689 and
CO velocity fields of NGC~4605 and NGC~5963.  In combination with our
previous study of NGC~2976, the full sample of five galaxies contains
density profiles that span the range from $\alphadm = 0$ to $\alphadm
= 1.20$, where $\alphadm$ is the power law index describing the
central density profile.  The scatter in $\alphadm$ from galaxy to
galaxy is 0.44, three times as large as in Cold Dark Matter (CDM)
simulations, and the mean density profile slope is $\alphadm = 0.73$,
shallower than that predicted by the simulations.  These results call
into question the hypothesis that all galaxies share a universal dark
matter density profile.  We show that one of the galaxies in our
sample, NGC~5963, has a cuspy density profile that closely resembles
those seen in CDM simulations, demonstrating that while galaxies with
the steep central density cusps predicted by CDM do exist, they are in
the minority.  In spite of these differences between observations and
simulations, the relatively cuspy density profiles we find do not
suggest that this problem represents a crisis for CDM.  Improving the
resolution of the simulations and incorporating additional physics may
resolve the remaining discrepancies.  We also find that four of the
galaxies contain detectable radial motions in the plane of the galaxy.
We investigate the hypothesis that these motions are caused by a
triaxial dark matter halo, and place lower limits on the ellipticity
of the orbits in the plane of the disk of 0.043-0.175.
\end{abstract}


\section{INTRODUCTION}

Over the last several years, one of the most persistent problems
confronting the Cold Dark Matter (CDM) model has been the dichotomy
between observed galaxy density profiles and those seen in
cosmological simulations.  The simulations generate dark matter halos
with central density cusps of $\rho \propto r^{-1}$ or steeper
\citep[e.g.,][] {nfw96,moore99}.  Observations of dwarf and low
surface-brightness (LSB) galaxies, though, have usually shown that
density profiles with shallow central cores, where the density is
nearly constant with radius, fit the data better than cuspy profiles
\citep{
moore94,burkert95,blo99,blo01,db01a,bs01,db01b,dbb02,swb03,wdbw03,s03}.
This disparity is important because density profiles represent some of
the strongest available constraints on the CDM simulations at small
spatial scales.

Because most observational errors tend to make density profiles look
shallower than they actually are, whether this disagreement is real or
only apparent has been a point of contention.  Recently, \citet{vdb00}
and \citet{vdbs01} put forth the argument that most existing rotation
curves have neither the spatial resolution nor the velocity resolution
necessary to test the simulations adequately. \citet{smvb03} extended
this argument with simulations showing that even high-resolution data
may be subject to numerous systematic effects that can make density
profiles appear artificially shallow.  However, \citet{dbb02} and
\citet{dbbm03} use observations and simulations to reach the opposite
conclusion: that systematic uncertainties do not significantly affect
most observed rotation curves.  In an attempt to resolve this debate,
we began an effort to measure the density profiles of a sample of
nearby dark matter-dominated galaxies at very high spatial and
velocity resolution, using improved techniques to minimize the
importance of systematic uncertainties.

In two previous papers, we reported on our results for the dwarf
spiral galaxies NGC~4605 \citep{us} and NGC~2976 \citep{s03}.  We
found that NGC~2976 contains a nearly constant density core, while
NGC~4605 has a density profile that is intermediate between the
usually observed constant density cores and the cusps predicted by
simulations.  In this paper, we present similar analyses of three more
nearby galaxies, NGC~5949, NGC~5963, and NGC~6689.  Since our earlier
study of NGC~4605 did not include a two-dimensional \ha\ velocity
field, which prevented us from confirming the existence of the radial
motions suggested by the CO data, we also update that work with an
\ha\ velocity field, additional CO mapping, and new optical imaging.
As before, we use high-resolution two-dimensional velocity fields to
constrain the overall mass distribution, and optical and near-infrared
imaging to model and remove the baryonic contribution to the rotation
curve.

In the following section, we describe our observations and data
reduction.  We briefly explain our methods of modeling stellar disks
and constructing tilted-ring models of the velocity fields, and refer
the reader to \citet{s03} for additional details.  In \S
\ref{results}, we present the results of our rotation curve analysis
and fit the rotation curves with various functional forms.  In \S
\ref{discussion} we discuss the most important implications of this
work.  We consider the effect on the derived density profiles of
relaxing the assumption that dark matter halos are spherical in \S
\ref{triaxialeffects} and we present our conclusions in \S
\ref{conclusions}.

\section{OBSERVATIONS, DATA REDUCTION, AND METHODOLOGY}
\label{observations}

The primary objective of this study is to derive very accurate
rotation curves (and hence density profiles) of low-mass,\footnote{We
refrain from using the term \emph{dwarf galaxy} here to avoid
confusion caused by the range of definitions of the term in the
literature.  We instead refer to our targets as \emph{low-mass}
galaxies, by which we mean that they are significantly less massive
than giant galaxies like the Milky Way and thus can be expected to be
dark matter-dominated.} dark matter-dominated spiral galaxies.  We use
two-dimensional velocity fields to measure the rotation curves in
order to avoid the systematic problems associated with long-slit
spectroscopy \citep{smvb03,s03}.  Where possible, we obtain velocity
fields at multiple wavelengths (e.g., \ha\ and CO) as a further guard
against systematic errors.  Modeling and removing the stellar disk is
an important step in the derivation of the dark matter density
profile, so we use multicolor imaging to obtain the best available
estimate of the stellar mass-to-light ratio.

Our target galaxies were originally selected from the CO survey of
nearby dwarf galaxies by Leroy et al. (in preparation) in the hope
that we could map the CO emission in the centers of the galaxies with
the BIMA interferometer.  Only three galaxies in the survey (NGC~2976,
NGC~4605, and NGC~5963) contained enough CO to produce useful
interferometer maps, so we also added two more galaxies that were
similar in mass (with rotation velocities of approximately 100 \kms),
distance (approximately 10 Mpc), inclination, and luminosity to the
CO-rich dwarfs.

\subsection{Target Galaxies and Observations}
\label{targets}
The properties of NGC~4605 are described by \citet{us}.  This late
type galaxy has an absolute magnitude of M$_{B} = -17.7$ and it is
located $4.26 \pm 0.64$ Mpc away (M. Pierce 2001, private
communication).  While NGC~4605 is classified as a barred galaxy
(SBc), detailed images do not reveal evidence for the presence
of a bar.

NGC~5949 is an Sbc galaxy of similar luminosity (M$_{B} = -18.2$) at a
larger distance.  Since this galaxy was included in the Spiral Field
I-band (SFI) Tully-Fisher survey, we can use the observed parameters
given by \citet{haynes99a,haynes99b}, corrected for Galactic and
internal extinction and for turbulent broadening of the \hi\ line, and
the Tully-Fisher Relation determined by \citet{g97} to calculate a
distance of $14.0 \pm 2.4$ Mpc.

NGC~5963 is an Sc galaxy associated with the NGC~5866 group
\citep{fouque92}.  The nearest large galaxy in the group is NGC~5907,
at a projected distance of 430 kpc, so it is unlikely that NGC~5963 is
currently interacting with its neighbors.  It has a heliocentric
recession velocity of $654 \pm 3.1$ \kms.  There are no photometric
distance determinations for this galaxy in the literature, so it is
reasonable to assume that it lies at the distance of 13 Mpc implied by
the Hubble flow\footnote{We use the \emph{Hubble Space Telescope} Key
Project value of $H_{0} = 72$ \kms\ Mpc$^{-1}$ \citep{keyproj} for the
Hubble constant.} after correcting for the motion of the Local Group
towards Virgo.  The uncertainty on this distance is probably of order
3 Mpc.  At 13 Mpc, NGC~5963 has an absolute magnitude of M$_{B} =
-17.8$.

NGC~6689 (also called NGC~6690) is listed as SBcd in NED and SBc in
LEDA, but it is classified as an unbarred Sc galaxy in the UGC
\citep{ugc}, and Sd in the Third Reference Catalog of Bright Galaxies
\citep[][hereafter RC3]{rc3}.  No bar is evident in any of our optical
or near-infrared images.  We conclude that the conflicting
classifications are due to the relatively high inclination of the
galaxy and that NGC~6689 does not contain a bar.  The galaxy is rather
similar to the others (but more inclined), with a distance of 11 Mpc
(again using the Virgocentric-flow-corrected velocity) and an absolute
magnitude of M$_{B} = -17.6$.

\subsubsection{\ha\ Observations and Reductions}
\label{hadata}

Our \ha\ observations were obtained on 2002 March 20-21, 2002 May
25-28, and 2003 April 15 at the 3.5 m WIYN telescope with the DensePak
fiber array.  See \citet{bsh98} and \citet{s03} for details about the
instrument and spectrograph setup, respectively.

We observed the galaxies at five to eleven positions, depending on the
spatial extent of their \ha\ emission.  These observations resulted in
350-650 independent velocity measurements for each galaxy.  Exposure
times ranged from 10 minutes to 60 minutes per position.  The angular
resolution of these data is 4\arcsec\ and the velocity resolution is
13 \kms.

The DensePak data were reduced as described by \citet{s03}.  The only
significant change in the data reduction resulted from the replacement
of the CuAr comparison lamp with a ThAr lamp in 2002 May.  The very
bright Th lines near the wavelength of \ha\ allowed us to improve the
accuracy of the wavelength solution by about a factor of two (to
$\sim0.2$ \kms).

\subsubsection{CO Observations and Reductions}
\label{codata}

Our \co\ \jone\ observations of NGC~4605 and NGC~5963 were acquired
using the B, C, and D configurations of the 10-element BIMA array
\citep{w96} between April 2001 and March 2002.  The total integration
time for each galaxy was $\sim80$ hours, much of which was in the most
extended (B) configuration.  The CO emission in NGC~4605 extends
beyond the BIMA primary beam diameter of $\sim100\arcsec$, so we
constructed a mosaic of observations made at several positions along
the major axis of the galaxy.  (The CO observations of the central
field used in this paper are the same as those of \citealt{us}, and
the other major axis fields are from the new mosaic.)  Because of the
greater distance of NGC~5963, the CO emission in that galaxy is much
more compact and fits easily inside the primary beam.  Observational
setup and data reduction were identical to that described by
\citet{s03}.  Beamsizes and sensitivities were $5\farcs 8 \times
5\farcs 1$ ($120\times 105$ pc) and 24 mJy beam$^{-1}$ for the
NGC~4605 central field, and $15\farcs 1 \times 13\farcs 4$ ($312
\times 277$ pc) and 70 mJy beam$^{-1}$ for the outlying fields of the
mosaic.  For NGC~5963, the beamsize was $5\farcs8 \times 5\farcs3$
($370\times 331$ pc) and the sensitivity was 31 mJy beam$^{-1}$.  We
detected CO in NGC~5949 and NGC~6689 with observations at the UASO 12
m telescope (Leroy et al., in preparation), but the emission was not
bright enough to map with BIMA.

\subsubsection{Optical and Near-IR Imaging and Reductions}

\begin{figure*}[!ht]
\figurenum{1}
\epsscale{1.0}
\plotone{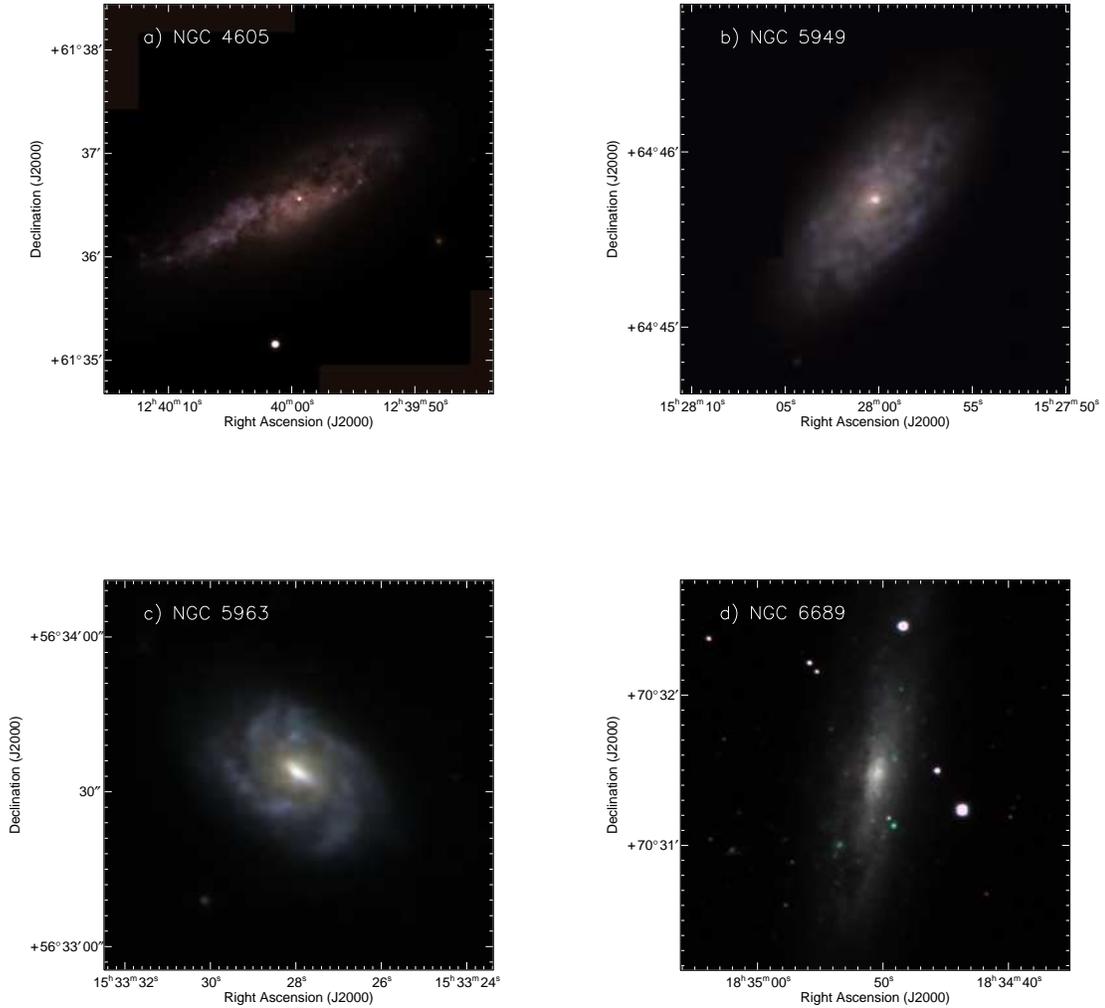}
\caption{Three-color images of our four target galaxies.  Each image
uses the asinh scaling recommended by \citet{lupton} to retain an
optimal combination of color and intensity information.  (a) $BVR$
image of NGC~4605 from Lowell Observatory. The southeast \twothirds\
of the galaxy is clearly brighter than the remainder.  (b) $BVR$ image
of NGC~5949 from Lick Observatory.  The blacked-out rectangle on the
east side of the galaxy is where a large dust grain was not completely
removed by flatfielding.  We masked out the affected area for the
photometric analysis.  (c) $BVR$ image of NGC~5963 from Lowell
Observatory. (d) $r^{\prime} \ha i^{\prime}$ image of NGC~6689 from
Kitt Peak National Observatory.
\label{colorims}}
\end{figure*}

We observed NGC~4605 and NGC~5963 with $B$, $V$, $R$, and $I$ filters
at the 1.8~m Perkins Telescope at Lowell Observatory on 2002 February
11.  NGC~5963 was also observed with much longer exposure times in $B$
and $R$ with the 1~m Nickel Telescope at Lick Observatory on 2003 June
23-24 in order to probe farther out into the LSB disk of the galaxy.
NGC~5949 was imaged in $B$, $V$, $R$, and $I$ at the Nickel on 2003
September 11, and NGC~6689 was observed in Sloan $r^{\prime}$ and
$i^{\prime}$ bands with the Mosaic camera on the 4~m Mayall Telescope
at Kitt Peak on 2003 October 20.  All imaging took place under
photometric conditions.  Reduction and photometric calibration of
these images followed the description of \citet{s03}.  To extend our
set of images to the near-infrared, we used the 2MASS $J$, $H$, and
$K_{\mbox{s}}$ Atlas images of each galaxy.  Three-color optical
images of all four galaxies are displayed in Figure \ref{colorims}.

\subsection{Isophotal fits and Stellar Disk Rotation Curves}
\label{stellardisks}

\begin{figure*}[!ht]
\figurenum{2}
\epsscale{1.0}
\plotone{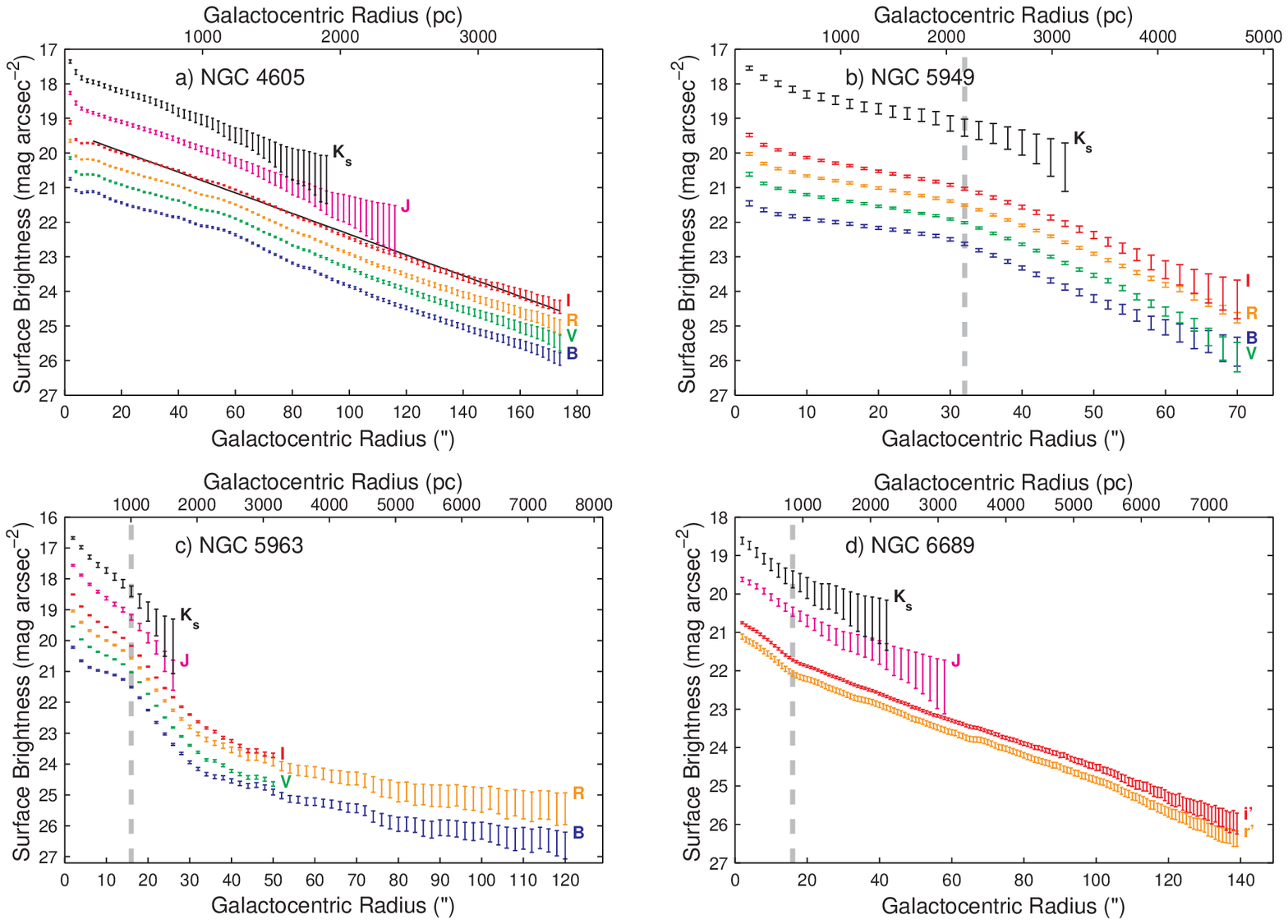}
\caption{(a) Optical and near-infrared surface brightness profiles of
NGC~4605.  In each band, the nucleus and exponential disk are visible.
The black solid line represents an exponential fit to the $I$-band
profile; the maximum deviation from the data is 0.13 magnitudes.  (b)
Optical and near-infrared surface brightness profiles of NGC~5949.  In
each band, the nucleus, exponential inner disk, and exponential outer
disk are all visible.  (c) Optical and near-infrared surface
brightness profiles of NGC~5963.  (d) Optical and near-infrared
surface brightness profiles of NGC~6689.  For all galaxies, the $J$ and
$K_{\mbox{s}}$ data can be traced further out, but we do not plot the
data beyond where the uncertainties reach a factor of two (0.75 mag).
The $H$-band profiles of each galaxy have been omitted for clarity; the
error bars for $H$ and $K_{\mbox{s}}$ overlap at most radii.  The
vertical dashed lines in panels b), c), and d) emphasize breakpoints
in the profiles.  The galaxies are generally well described by
exponential disks, except for NGC~5963, which has an unusual and
difficult to characterize surface brightness profile.  We speculate
that this galaxy may have evolved a dense, bulgelike central region
through secular processes (a pseudobulge).
\label{sbprofs}}
\end{figure*}

We used the optical/near-IR images of these galaxies to place
reasonable limits on the contributions of their stellar disks to their
rotation curves.  We extracted surface brightness profiles from the
images with the IRAF task {\sc ellipse}, as described in \citet{s03}.
For each image of each galaxy, we ran {\sc ellipse} allowing the
center position, position angle (PA), and ellipticity of the isophotes
to vary with radius.  We then defined the isophotal center of the
galaxy in that band to be the weighted average of the ellipse centers.
{\sc Ellipse} was then run again with the center fixed, and we
measured the weighted averages of the PA and ellipticity.  We then
averaged the isophotal centers\footnote{For the two galaxies with
visually obvious nuclei (NGC~4605 and NGC~5949), we found that the
isophotal centers were located quite close to the nuclei.  In these
cases, we fixed the ellipse centers on the nuclei instead of the
formal isophotal centers.}, position angles, and ellipticities for
each galaxy in every available band to define fiducial values.  In
cases where a trend as a function of wavelength was noticed for any of
the parameters, values for the redder bands were given preference to
minimize the importance of extinction.  Finally, we ran {\sc ellipse}
one more time with all three parameters fixed in order to determine
the final surface brightness profiles, which are shown in Figure
\ref{sbprofs}.  The measured isophotal parameters for each galaxy are
listed in Table \ref{isophotaldata}.  Discussion of the various
surface brightness profiles can be found in \S \ref{comments}.

\begin{deluxetable*}{c c c c c c}
\tablewidth{0pt}
\tablecolumns{6}
\tablecaption{Measured Isophotal Parameters of Target Galaxies}
\tablehead{
\colhead{Galaxy} & \colhead{$\alpha$ (J2000.0)} & \colhead{$\delta$
(J2000.0)} & \colhead{PA} & \colhead{ellipticity} &
\colhead{inclination} }
\startdata
NGC~4605 &  12\hr39\min59\fs 27  & 61\degr36\arcmin33\farcs3  &  119\degr  &  0.630  & 71.5\degr \\
NGC~5949 &  15\hr28\min00\fs 69  & 64\degr45\arcmin47\farcs7  &  144\degr  &  0.535  & 64.6\degr \\
NGC~5963 &  15\hr33\min27\fs 90  & 56\degr33\arcmin35\farcs0  &  \phn54\degr  &  0.320  & 48.4\degr \\
NGC~6689 &  18\hr34\min50\fs 18  & 70\degr31\arcmin27\farcs1  &  170\degr  &  0.690  & 76.0\degr \\
\enddata
\label{isophotaldata}
\end{deluxetable*}

\citet{bdj01} showed that the color and stellar mass-to-light ratio
($\msl$) of a galaxy are correlated.  In order to obtain an estimate
at the mass-to-light ratios of our targets, we used the observed
colors and the updated color-M/L relations given by \citet{bell03} to
compute expected $R$ band mass-to-light ratios.  For NGC~6689, since
we do not have any optical observations in the Johnson-Cousins system,
we instead used the relations \citet{bell03} defined for the Sloan
magnitude system.  Measured colors and mass-to-light ratios are given
in Table \ref{mltable}.  Two apparent flaws in the \citet{bell03}
models are evident from our calculations: 1) the $B-V$ and $B-R$
colors do not predict consistent $R$ band mass-to-light ratios, and 2)
the predicted $K$ band mass-to-light ratios appear to be significantly
too high.

\begin{deluxetable*}{l c c c}
\tablewidth{0pt}
\tablecolumns{4}
\tablecaption{Stellar Mass-to-Light Ratios}
\tablehead{
\colhead{} & \colhead{Mean Inner} &
\colhead{\mslr} & \colhead{\mslk} \\
\colhead{} & \colhead{Disk Color\tablenotemark{a}} &
\colhead{[\mlr]} & \colhead{[\mlk]} \\
}

\startdata
\emph{NGC 4605}  &                         &        &       \\
Predictions from $B-V$ color\tablenotemark{b} &  0.45\tablenotemark{c}  &  0.94  &  0.72 \\
Predictions from $B-R$ color\tablenotemark{b} &  0.82\tablenotemark{c}  &  1.09  &  0.71 \\
Observed maximum disk values &                         &  1.36  &  0.37 \\
\\[2mm]
\emph{NGC 5949}  &                         &        &       \\
Predictions from $B-V$ color\tablenotemark{b} &  0.63\tablenotemark{d}  &  1.48  &  0.76 \\
Predictions from $B-R$ color\tablenotemark{b} &  1.14\tablenotemark{d}  &  1.80  &  0.78 \\
Observed maximum disk values &                         &  2.07  &  0.49 \\
\\[2mm]
\emph{NGC 5963}  &                         &        &       \\
Predictions from $B-V$ color\tablenotemark{b} &  0.51\tablenotemark{e}  &  1.09  &  0.73 \\
Predictions from $B-R$ color\tablenotemark{b} &  0.97\tablenotemark{e}  &  1.38  &  0.74 \\
Observed maximum disk values &                         &  2.09  &  0.61 \\
\\[2mm]
\emph{NGC 6689}  &                         &        &       \\
Predictions from $r^{\prime}-i^{\prime}$ color\tablenotemark{b} &  0.22\tablenotemark{f} &  1.96\tablenotemark{g} & 0.78 \\
Observed maximum disk values                    &                        &  3.14  &  1.07 \\
\enddata

\tablenotetext{a}{These colors have been corrected for
Galactic extinction and internal extinction.  The Galactic extinction
corrections were taken from \citet*{sfd98} and the internal extinction
corrections were derived with the method of \citet{sakai00} using the
axis ratios from our isophotal fits and \hi\ velocity widths extracted
from LEDA.}
\tablenotetext{b}{Calculated from the relations given in Appendix A of
\citet{bell03}.}
\tablenotetext{c}{NGC~4605 colors were measured for $10\arcsec \le r
\le 140\arcsec$.}
\tablenotetext{d}{NGC~5949 colors were measured for $6\arcsec \le r
\le 32\arcsec$.}
\tablenotetext{e}{NGC~5963 colors were measured for $8\arcsec \le r
\le 32\arcsec$.}
\tablenotetext{f}{NGC~6689 colors were measured for $16\arcsec \le r
\le 138\arcsec$.}
\tablenotetext{g}{Note that this mass-to-light ratio is actually
calculated for the Sloan $r^{\prime}$ band, not the Kron-Cousins $R$
band.}
\label{mltable}
\end{deluxetable*}

Stellar rotation curves were calculated via numerical integration of
the stellar surface mass densities (derived from the observed surface
brightness profiles) using the NEMO software package \citep{teuben}.
The primary assumption required by the NEMO implementation of this
method is that the disks are infinitesimally thin.  Removing this
assumption leaves the shape of the stellar rotation curve unchanged
but modestly lowers its amplitude.

\subsection{Velocity Field Fitting}
\label{velfieldfitting}

\begin{figure*} [!ht]
\figurenum{3}
\epsscale{1.0}
\plotone{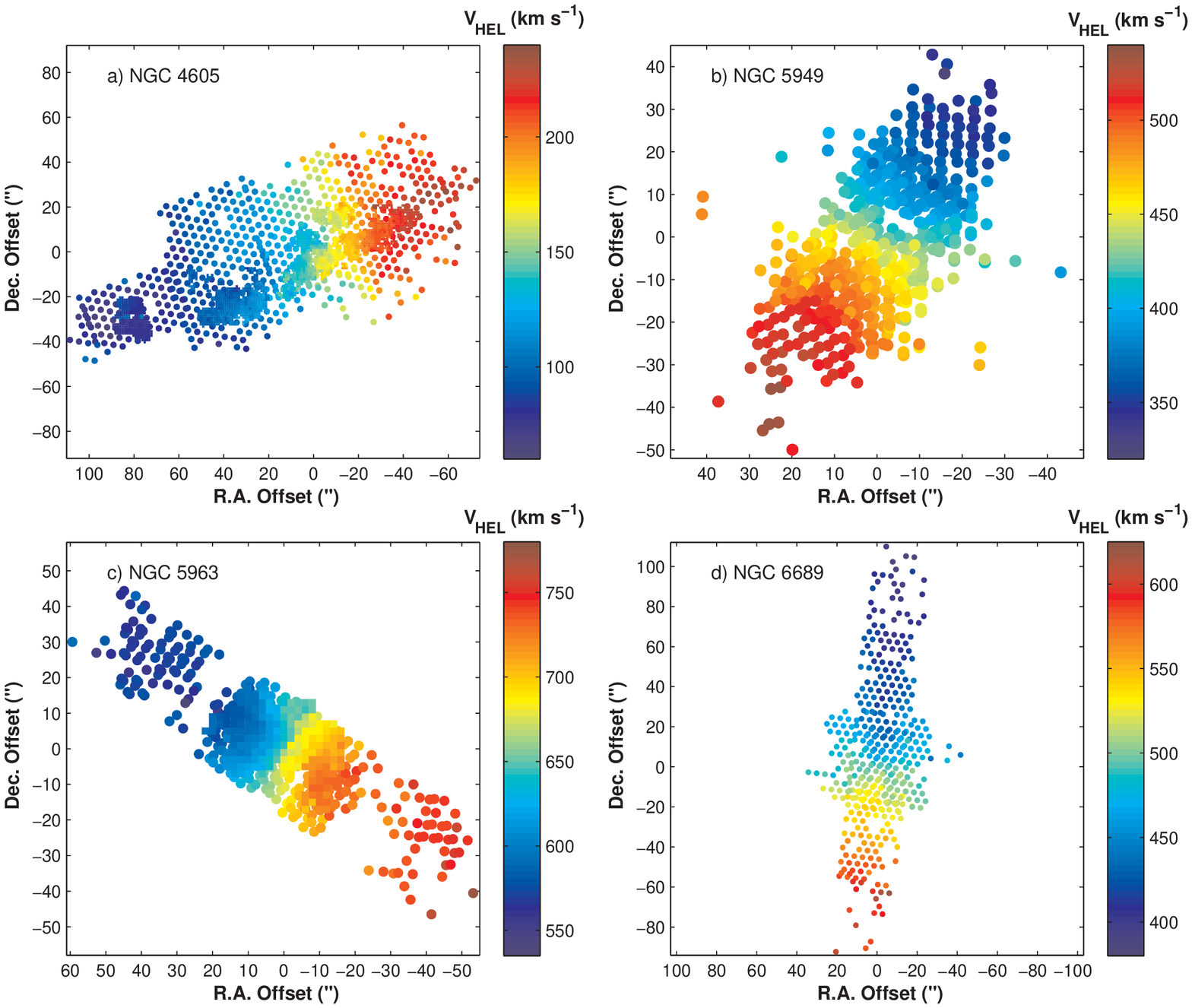}
\caption{Velocity fields of the four galaxies discussed in this paper.
(a) Combined \ha\ (circles) and CO (filled-in) velocity field of
NGC~4605.  (b) \ha\ velocity field of NGC~5949.  (c) Combined \ha\ and
CO velocity field of NGC~5963.  (d) \ha\ velocity field of NGC~6689.
\label{vfields}}
\end{figure*}

The velocity fields of the four galaxies are shown in Figure
\ref{vfields}.  The circles represent fiber-based \ha\ measurements,
and the filled-in regions show CO measurements.  In general, the
velocity fields appear quite regular, with only NGC~4605 showing any
sign that its kinematic and photometric minor axes might not be
aligned.

\subsubsection{Comparison of \ha\ and CO Data}
\label{havsco}

For NGC~4605 and NGC~5963, we have resolved CO velocity fields in
addition to the \ha\ observations.  To check for agreement between
these different data sets, we compared the \ha\ and CO velocities at
every position where emission is visible from both species.  In the
case of NGC~5963, we find a mean offset of less than 1 \kms, with a
scatter of 7.8 \kms, indicating excellent agreement between the \ha\
and CO velocities.  For NGC~4605, only the westernmost molecular cloud
has a velocity that agrees exactly with the velocity of the
surrounding ionized gas.  Elsewhere in the galaxy there is a small but
consistent offset of 4.8 \kms, in the sense that the \ha\ velocities
are smaller.  The scatter between the two velocity fields is 8.4 \kms.
Nevertheless, tilted-ring models based only on the CO velocity field
match \ha-only tilted-ring models quite closely, so this velocity
offset does not affect our results.  Velocity differences of this
magnitude are expected to arise from flows of the ionized gas away
from molecular clouds \citep{ftb82,fdt90}.  Since the \ha\ and CO data
appear to be both reliable and consistent with each other, the
velocity field fitting described in the following section was
performed on the combined velocity data for these two galaxies.

\subsubsection{Tilted-Ring Modeling and Error Analysis}
\label{ringfit}

We derived rotation curves for each galaxy in the same way as in
\citet{s03}.  Using the geometric parameters measured from the
photometry (and listed in Table 1), we ran the tilted-ring modeling
routine \ringfit\ to extract the rotational, radial, and systemic
velocities as a function of radius from the velocity fields.  As
detailed in \citet{s03}, \ringfit\ is similar to the standard \rotcur\
algorithm \citep{b87} in that it includes both rotation and radial
motions in its ring fits, but it does not allow the PA, inclination
angle, or center to vary from ring to ring.  Next, we applied a
bootstrap technique to estimate the uncertainties in the PA, center
position, and inclination angle.  Typical bootstrap uncertainties were
1\arcsec\ in RA and Dec, $4\degr$ in PA, and $2\degr$ in inclination.
For all four galaxies, we found that the kinematic and photometric
values of the PA, center, and inclination angle agreed within the
errors.

After running the bootstrap, we constructed Gaussian distributions of
each of the geometric parameters, assuming that the full width at half
maximum of the Gaussian was equal to the bootstrap uncertainty.  We
then created 1000 Monte Carlo realizations of each rotation curve by
randomly selecting a PA, center position, and inclination angle from
the Gaussian distributions and running \ringfit\ with the selected
parameters as inputs.  This process resulted in 1000 Monte Carlo
rotation velocities at each radius.  We then defined our final
rotation curves to be the means of the Monte Carlo rotation
velocities, and the uncertainties on the rotation curves to be the
standard deviations of the Monte Carlo rotation velocities.  The final
radial velocity curves were derived in the same manner.  By
incorporating the uncertainties of the PA, inclination, and center
position, this technique yields more realistic rotation curve error
bars than simply propagating the very small velocity uncertainties
from each spectrum through the analysis.  The Monte Carlo
uncertainties (which we refer to as systematic uncertainties) are
always much larger than the statistical uncertainties calculated by
standard error propagation.  The uncertainties we use for the
remainder of the paper are the sum in quadrature of the systematic and
statistical uncertainties.

\subsubsection{Higher Order Harmonic Fits}
\label{harmonicfits}

In addition to the standard use of \ringfit\ described in the previous
section, the algorithm can also be used to decompose the velocity
field into Fourier components up to third order ($\sin{3\theta}$ and
$\cos{3\theta}$).  If present, these higher order terms strongly
suggest the existence of a bar or other non-axisymmetric perturbation
to the gravitational potential.  With enough S/N, the unique
signatures of a bar, spiral arms, an elliptical potential, or
lopsidedness can be detected \citep{schoen97}.  We find weak evidence
for nonzero higher order components in the velocity fields of
NGC~4605, NGC~5949, and NGC~5963.  Each galaxy has several adjacent
rings in which the fits deviate from zero by $\sim2\sigma$ (all four
higher-order terms for NGC~4605, $\cos{2\theta}$ and $\sin{3\theta}$
for NGC~5949, and $\cos{2\theta}$ and $\sin{2\theta}$ for NGC~5963).
Typical amplitudes of the higher order components are $2-5$ \kms,
which should not be large enough to affect our derivation of the
density profiles in the following section.  Although the significance
of these components does not generally exceed 2.5$\sigma$ in a given
ring, we argue that these detections are probably real because 1) a
number of consecutive rings have consistent Fourier components, and 2)
the rotation curve uncertainties appear to be overestimated (see \S
\ref{otherfits}).  NGC~6689 does not have enough data points in most
of its rings to obtain reliable higher-order fits.  For comparison, we
find $3-5$ \kms\ $\cos{2\theta}$, $\cos{3\theta}$, and $\sin{3\theta}$
components over small ranges of radii in the NGC~2976 velocity field
studied by \citet{s03}.  Including terms up to third order in the
tilted-ring models (seven free \ringfit\ parameters instead of three)
does significantly improve the reduced $\chi^{2}$ value of the fits in
many cases, but this is often because the number of data points in the
ring is not much larger than seven so that the fit can go through
every point.  More detailed descriptions of the higher order motions
in each galaxy can be found in \S \ref{comments}.

\section{Results}
\label{results}

In this section we fit the rotation curves with a variety of
functional forms to ascertain the shapes of the density profiles.  We
first consider the likely contribution of the stellar disk to the
rotation curve, and then remove it from the data to reveal the
rotation curve of the dark matter halo.

\begin{figure*}[!ht]
\figurenum{4}
\epsscale{1.0}
\plotone{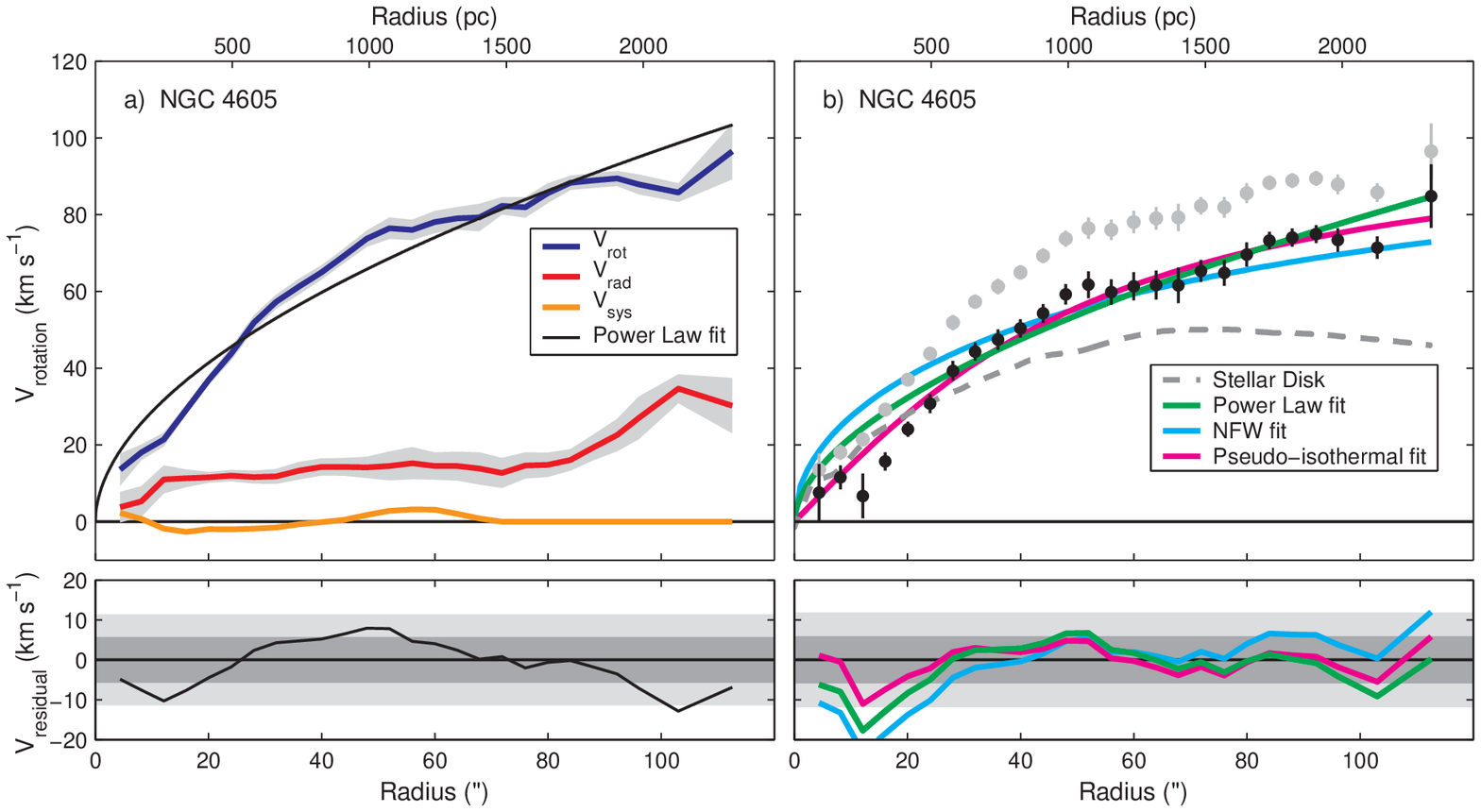}
\caption{(a) Tilted-ring model for NGC~4605.  The thick blue curve
represents the rotation curve, and the thick red and orange curves
show the radial velocity curve and the systemic velocities,
respectively.  Starting at a radius of 72\arcsec\ the systemic
velocities were held fixed at zero.  The shaded gray regions
surrounding the rotation and radial velocity curves represent the
combined systematic and statistical $1\sigma$ uncertainties from the
bootstrap and Monte Carlo analysis.  Note that for convenience we have
plotted the radial velocities as positive; this choice does not
indicate whether the radial motions are directed inwards or outwards.
The thin black curve is a power law fit to the rotation curve.  The
residuals from this fit are plotted in the lower panel, and the
$1\sigma$ and $2\sigma$ scatter of the data points around the fit is
shown by the shaded gray areas.  (b) Disk-subtracted rotation curve of
NGC~4605.  The gray points are the original rotation velocities from
panel (a), and the black points are the dark matter halo rotation
curve after subtracting the stellar disk ($\mslr = 1.01\mlr$), which
is shown as a gray dashed line.  The thick green, cyan, and magenta
curves show power law, NFW, and pseudo-isothermal fits to the halo
rotation curve, respectively.  The residuals from these fits are
displayed in the lower panel, and the $1\sigma$ and $2\sigma$ scatter
of the data points around the power law fit is shown by the shaded
gray areas.
\label{rcfig4605}}
\end{figure*}

\begin{figure*}[!ht]
\figurenum{5}
\epsscale{1.0}
\plotone{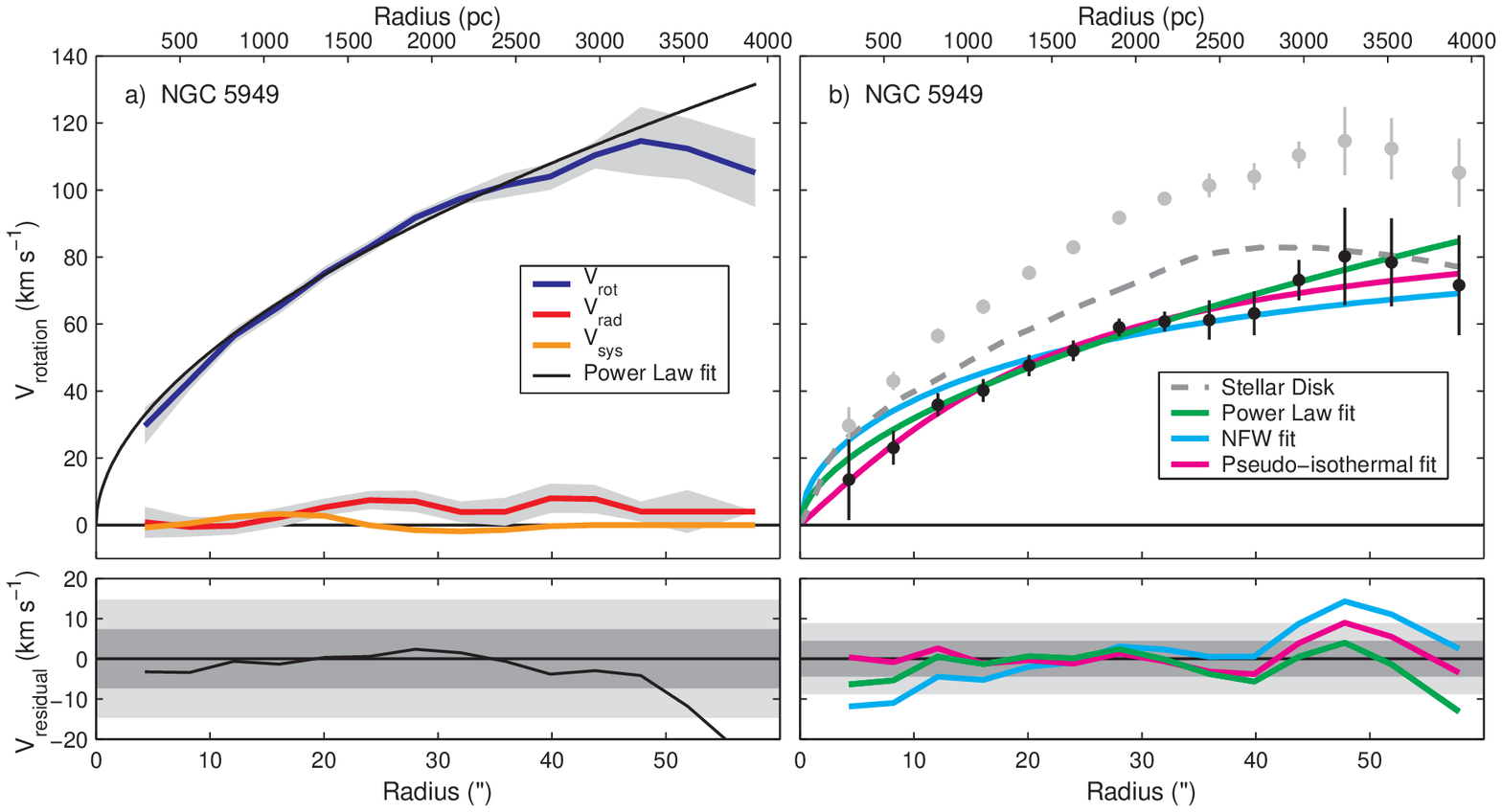}
\caption{(a) Tilted-ring model for NGC~5949.  (b) Disk-subtracted
rotation curve of NGC~5949 (for $\mslr = 1.64\mlr$).  Symbols, colors,
and shading are as in Figure \ref{rcfig4605}.
\label{rcfig5949}}
\end{figure*}

\begin{figure*}[!ht]
\figurenum{6} 
\epsscale{1.0} 
\plotone{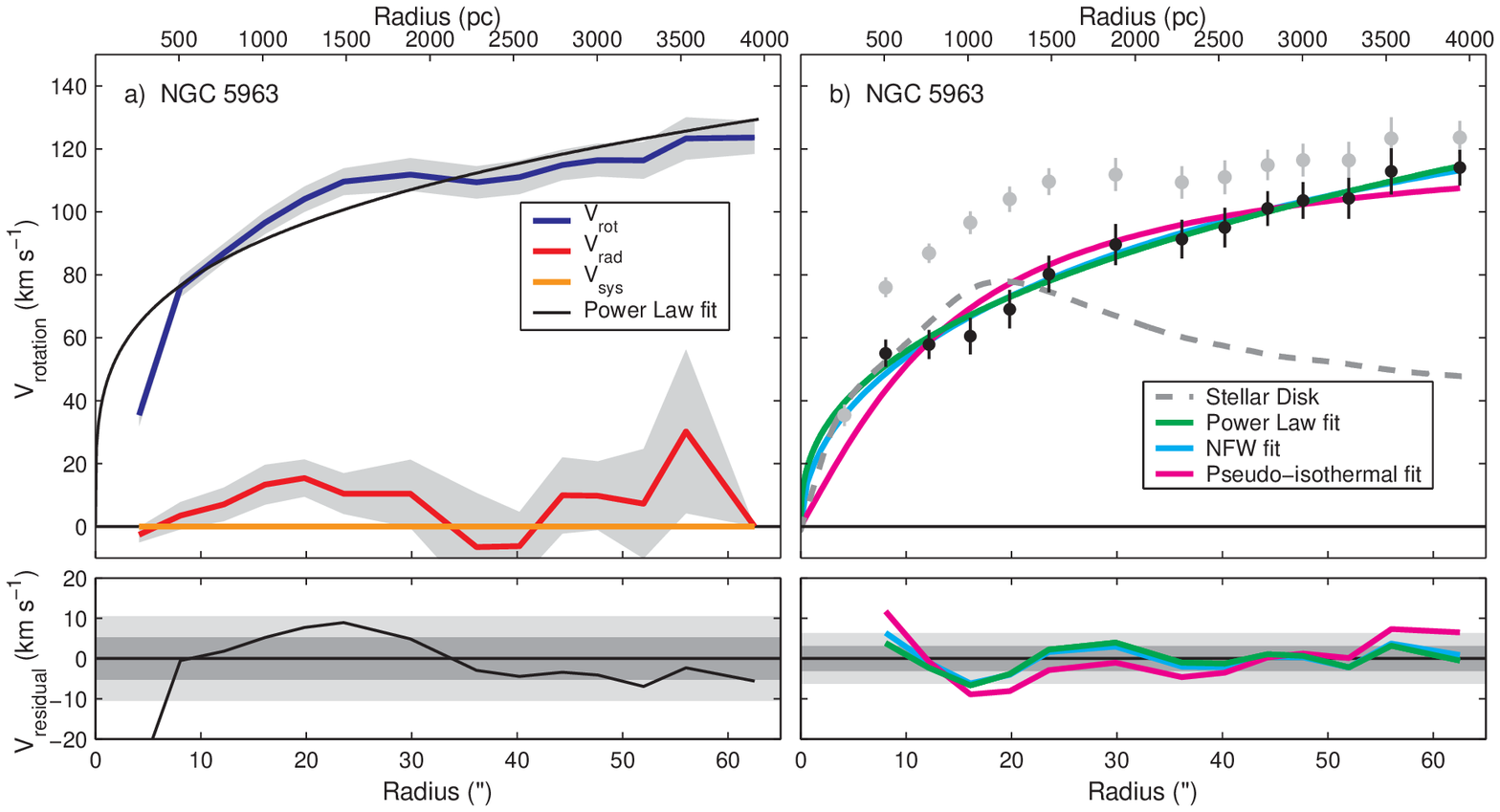}
\caption{(a) Tilted-ring model for NGC~5963.  (b) Disk-subtracted
rotation curve of NGC~5963 (for $\mslr = 1.24\mlr$).  Symbols, colors,
and shading are as in Figure \ref{rcfig4605}.
\label{rcfig5963}}
\end{figure*}

\begin{figure*}[!ht]
\figurenum{7}
\epsscale{1.0}
\plotone{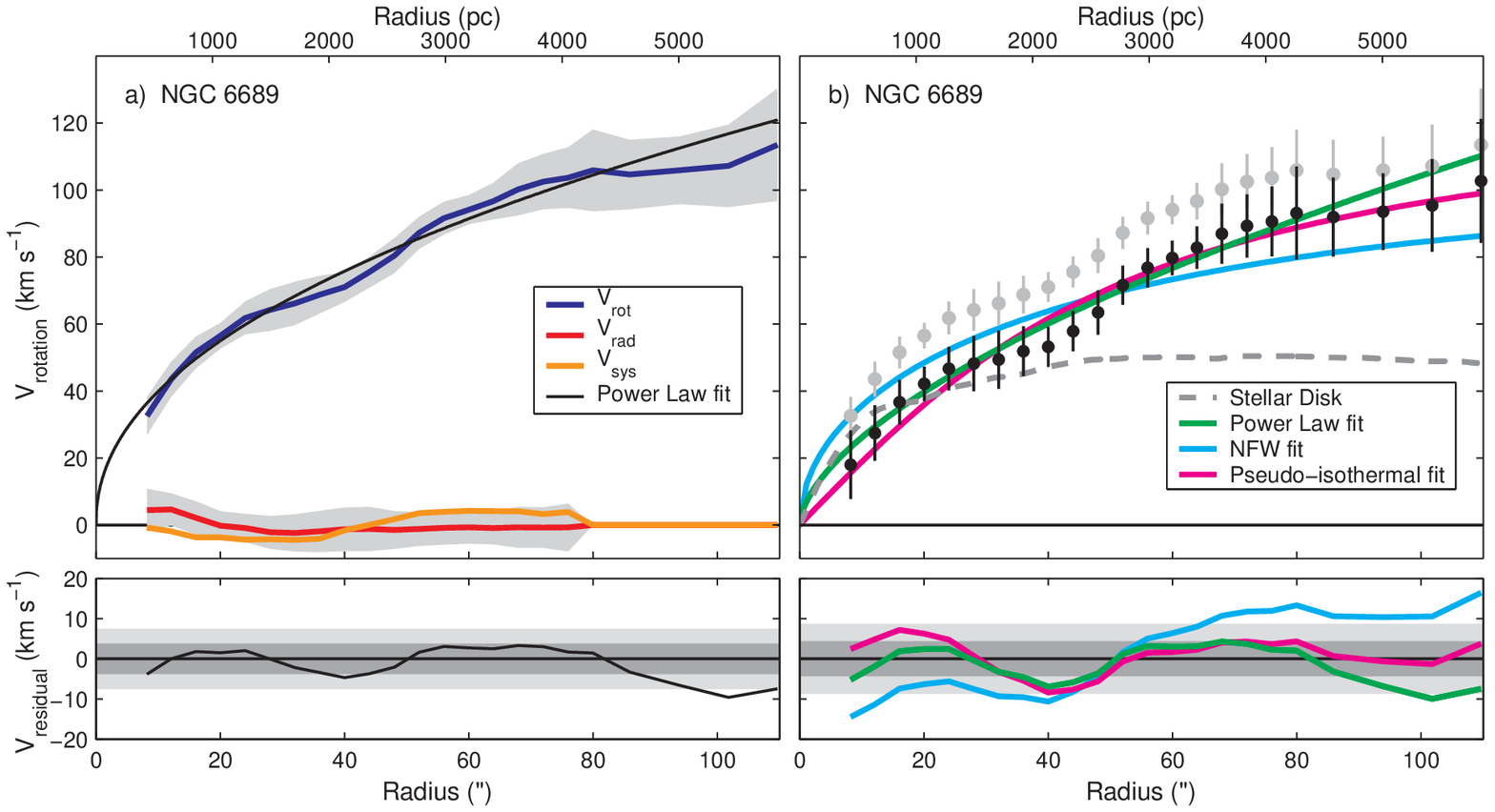}
\caption{(a) Tilted-ring model for NGC~6689.  (b) Disk-subtracted
rotation curve of NGC~6689 (for $\mslrs = 1.96\mlrs$).  Symbols,
colors, and shading are as in Figure \ref{rcfig4605}.
\label{rcfig6689}}
\end{figure*}

The rotation curves produced by the tilted-ring models in \S
\ref{ringfit} are displayed in Figures \ref{rcfig4605}a -
\ref{rcfig6689}a.  Radial velocity curves and the uncertainties on
both the rotation velocities and the radial velocities are also
plotted in the same figures.

\subsection{Removing the Baryons}
\label{maxdisk}

What we are actually interested in are the rotation curves of the
\emph{dark matter halos} of these galaxies, which are only equivalent
to the total rotation curves shown in Figures \ref{rcfig4605}a -
\ref{rcfig6689}a in the minimum disk scenario ($\msl = 0$).  We
therefore need to account for the contribution of the baryons to the
rotation curves.  An \emph{upper limit} on the dark matter rotation
curve (and also the slope of the density profile) can be found if the
disk mass is zero, and a \emph{lower limit} to the dark matter
rotation curve and density profile slope is obtained for a maximum
disk.  In general, for galaxies of normal surface brightness, the
minimum disk solution is physically unrealistic and the actual mass
distribution is likely to be closer to the maximum disk case.

We used the stellar disk rotation curves calculated in \S
\ref{stellardisks} to construct maximum disk models\footnote{ With the
possible exception of NGC~6689, these galaxies do not appear to
contain bulges, so the stellar disk is the only baryonic component
that we model. Including the gas disk as well would increase the
effective mass-to-light ratio of the disk somewhat, but does not
affect our results (see following paragraph).}.  In order to avoid
being unduly influenced by a single unrepresentative point in any of
the observed rotation curves, we fit the inner five points of each
rotation curve with a smooth function (a power law).  We then
incrementally increased the stellar mass-to-light ratio (the stellar
rotation curve scales as $v_{*,rot} \propto \sqrt{\msl}$) until the
stellar rotation curve exceeded the fit to the observed points.
Because of the influence of the nucleus on the stellar rotation curve
of NGC~4605, we ignored the innermost point of the fit to the observed
rotation curve for that galaxy.  With this procedure we defined the
following maximum disk mass-to-light ratios: $\mslr = 1.36 \mlr$
(NGC~4605), $\mslr = 2.07 \mlr$ (NGC~5949), $\mslr = 2.09 \mlr$
(NGC~5963), and $\mslrs = 3.14 \mlrs$ (NGC~6689).  Typical
uncertainties on these values are $\sim 15$ \%.  For all of the
galaxies, these kinematic mass-to-light ratios are significantly
higher than the mass-to-light ratios predicted from the galaxy colors
by the \citet{bell03} population synthesis models (see Table
\ref{mltable}).  We conclude the disks of these galaxies are probably
somewhat submaximal.  For the remainder of this paper, we adopt the
following mass-to-light ratios (the average predicted from the $B-V$
and $B-R$ colors shown in Table \ref{mltable}): $\mslr = 1.01 \mlr$
(NGC~4605), $\mslr = 1.64 \mlr$ (NGC~5949), $\mslr = 1.24 \mlr$
(NGC~5963), and $\mslrs = 1.96 \mlrs$ (NGC~6689).

Note that since we do not have \hi\ data, we are neglecting the
contribution of the gas to the rotation curve in this analysis.  In
galaxies the size of our targets, however, the stellar disk almost
always contributes significantly more mass at the center of the galaxy
than the gas does and the effect of including the gas is similar to a
$20 \%$ change in the stellar mass-to-light ratio \citep{us,s03}.  If
we arbitrarily increase \msl\ to simulate this effect, the slope of
the dark matter density profile (see below) decreases by $2-12$\%.
Allowing the stellar disk to have a nonzero thickness (see \S
\ref{stellardisks}) would offset this decrease.

\subsection{Power Law Density Profile Fits}
\label{powerlaw}

We can now consider the central density profile slopes of the dark
matter halos of each galaxy.  Following the notation of \citet{s03},
we parameterize the rotation curve by $v_{rot} \propto
r^{(2-\alphadm)/2}$, which corresponds to a power law density profile:
$\rhodm \propto r^{-\alphadm}$.  Allowing the stellar mass-to-light
ratios to vary from zero up to the maximum disk values determined in
the previous paragraph, the limits we place on $\alphadm$ are given in
Table \ref{alphalimits}.  Obviously, the impact of changing the
mass-to-light ratio on the density profiles is minimal for NGC~4605
and NGC~5949.  These galaxies have very similar dark matter density
profiles in the maximum and minimum disk cases, so their density
profile slopes are tightly constrained.

\begin{deluxetable}{c c c}
\tablewidth{0pt}
\tablecolumns{3}
\tablecaption{Limits on Dark Matter Density Profile Slopes}
\tablehead{
\colhead{Galaxy} & \colhead{maximum disk} & \colhead{minimum disk} \\
\colhead{} & \colhead{$\alphadm$} & \colhead{$\alphadm$} 
}

\startdata
NGC~2976 & $0.01 \pm 0.13$  &  $0.27 \pm 0.09$   \\
NGC~4605 & $0.71 \pm 0.06$  &  $0.90 \pm 0.02$   \\
NGC~5949 & $0.79 \pm 0.17$  &  $0.93 \pm 0.04$   \\
NGC~5963 & $0.75 \pm 0.10$  &  $1.41 \pm 0.03$  \\
NGC~6689 & $0.43 \pm 0.18$  &  $1.07 \pm 0.06$   \\
\enddata
\label{alphalimits}
\end{deluxetable}

The rotation curve fits for these limiting cases are displayed in
Figures \ref{rcfig4605} - \ref{rcfig6689}.  In the left panel of each
figure, a minimum disk power law fit is displayed, with the residuals
from the fit in the lower panel.  The right panel of each figure
contains a power law fit to the dark matter halo in the maximum disk
case, again with residuals shown in the lower panel.  In all cases a
power law provides a reasonable fit to the rotation curve at least out
to the radius at which the rotation curve begins to flatten
significantly (at which point the density profile slope is obviously
changing with radius and a single power law cannot be expected to
accurately represent the rotation curve).  The numerical parameters of
the power law fits for the disk-subtracted rotation curves are listed
in Table \ref{fitdata1}, and for the total rotation curves (with no
disk subtraction) in Table \ref{fitdata2}.  The dark matter density
profile slopes we derive after subtracting the stellar disks are
$\alphadm = [0.78, \mbox{ }0.88, \mbox{ }1.20, \mbox{ }0.79]$ for
NGC~4605, NGC~5949, NGC~5963, and NGC~6689, respectively.  If we fit
the velocity fields without allowing radial motions, we find density
profile slopes of $\alphadm = [0.64, \mbox{ }0.98, \mbox{ }1.19,
\mbox{ }0.79]$.  These changes are small (comparable to or less than
the $1\sigma$ uncertainties) and go in different directions for
different galaxies.  We conclude that including radial motions in our
tilted-ring models does not systematically affect the density profile
slopes we derive.
   
\begin{deluxetable*}{c c c c c c c c c c c c c}
\tablewidth{0pt}
\tablecolumns{13}
\tablecaption{Rotation Curve Fit Results After Disk Subtraction}
\tablehead{
  \multicolumn{1}{c}{} & \multicolumn{2}{c}{Power law}  & 
  \multicolumn{4}{c}{NFW} & \multicolumn{2}{c}{Pseudo-isothermal} &
  \multicolumn{4}{c}{} \\
  \multicolumn{1}{c}{} & \multicolumn{2}{c}{fit parameters}  & 
  \multicolumn{4}{c}{fit parameters}  & \multicolumn{2}{c}{fit parameters} &
  \multicolumn{4}{c}{} \\
  \multicolumn{1}{c}{} & 
  \multicolumn{2}{c}{$\overbrace{\phm{SpanningSpann}}$} &
  \multicolumn{4}{c}{$\overbrace{\phm{SpanningSpanningSpanningSpann}}$} & 
  \multicolumn{2}{c}{$\overbrace{\phm{SpanningSpanning}}$} & 
  \multicolumn{4}{c}{} \\
\colhead{Galaxy} & \colhead{$V_{0}$} & \colhead{$\alpha$} & 
\colhead{$r_{s}$} & \colhead{$R_{200}$} & \colhead{c} & 
\colhead{$V_{200}$} & \colhead{$\rho_{0}$} & \colhead{$R_{c}$} & 
\colhead{$\chi^{2}_{pl}$\tablenotemark{d}} & \colhead{$\chi^{2}_{NFW}$\tablenotemark{d}} & 
\colhead{$\chi^{2}_{iso}$\tablenotemark{d}} & \colhead{$\chi^{2}_{mod\_iso}$\tablenotemark{d}} \\
\colhead{} & \colhead{[\kms]} & \colhead{} & \colhead{[kpc]} & 
\colhead{[kpc]} & \colhead{} & \colhead{[\kms]} & 
\colhead{[\gpercmcu]} & \colhead{[kpc]} & \colhead{} & \colhead{} & 
\colhead{} & \colhead{} } 

\startdata
NGC~2976\tablenotemark{a} & 43.0 & 0.01 & \nodata\tablenotemark{b} & \nodata\tablenotemark{b} & \nodata\tablenotemark{b} & \nodata\tablenotemark{b} & $7.7 \times 10^{-24}$ & 2.5 & 0.27 & $>5.8$ & 0.43 & 1.00 \\
NGC~4605\phm{$^{a}$} & 51.4 & 0.78 & $>5.00$       & $>91$ & $<18.3$ & $>65$ & $2.0 \times 10^{-23}$ & 0.8 & 3.74 & $<8.03$ & 1.53 & 2.88 \\
NGC~5949\phm{$^{a}$} & 39.3 & 0.88 & $>6.40$       & $>87$ & $<13.6$ & $>62$ & $7.8 \times 10^{-24}$ & 1.2 & 0.38 & $<1.24$ & 0.24 & 0.27 \\
NGC~5963\phm{$^{a}$} & 66.4 & 1.20 & $\phm{>}10.8$ & $\phm{>}160.9$ & $\phm{>}14.9$    & \phn114.3 & $3.4 \times 10^{-23}$ & 0.8 & 0.34 & $\phm{>}0.43$ & 1.36 & 0.64 \\
NGC~5963\tablenotemark{c} & 74.5 & 1.50 & $\phm{>}6.71$ & $\phm{>}127.7$ & $\phm{>}19.0$    & \phn90.7 & $2.7 \times 10^{-23}$ & 0.9 & 2.19 & $\phm{>}0.42$ & 0.98 & 0.57 \\
NGC~6689\phm{$^{a}$} & 37.8 & 0.79 & $>8.7$       & $>108.8$ & $<12.5$ & $>77$ & $4.8 \times 10^{-24}$ & 2.1 & 0.30 & $<1.31$ & 0.46 & 0.30 \\

\enddata

\tablenotetext{a}{Using data from \citet{s03}.}
\tablenotetext{b}{No meaningful NFW fits could be made for NGC~2976,
even with only one parameter free.  The ``best'' fit had $\chi^2_{red} = 5.8$.}
\tablenotetext{c}{Using the rotation curve presented in this paper
plus the rotation velocities measured by \citet*{bosma88} at
$55\arcsec \le r \le 165\arcsec$.}
\tablenotetext{d}{All $\chi^{2}$ values listed here actually represent
the reduced $\chi^{2}$ ($\chi^{2}$ per degree of freedom in the fits).}

\label{fitdata1}
\end{deluxetable*}

\begin{deluxetable*}{c c c c c c c c c c c c c}
\tablewidth{0pt}
\tablecolumns{13}
\tablecaption{Rotation Curve Fit Results With No Disk Subtraction}
\tablehead{
  \multicolumn{1}{c}{} & \multicolumn{2}{c}{Power law}  & 
  \multicolumn{4}{c}{NFW}  & \multicolumn{2}{c}{Pseudo-isothermal} &
  \multicolumn{4}{c}{} \\
  \multicolumn{1}{c}{} & \multicolumn{2}{c}{fit parameters}  & 
  \multicolumn{4}{c}{fit parameters}  & \multicolumn{2}{c}{fit parameters} &
  \multicolumn{4}{c}{} \\
  \multicolumn{1}{c}{} & 
  \multicolumn{2}{c}{$\overbrace{\phm{SpanningSpann}}$} &
  \multicolumn{4}{c}{$\overbrace{\phm{SpanningSpanningSpanningSpann}}$} & 
  \multicolumn{2}{c}{$\overbrace{\phm{SpanningSpanning}}$} & 
  \multicolumn{4}{c}{} \\
\colhead{Galaxy} & \colhead{$V_{0}$} & \colhead{$\alpha$} & 
\colhead{$r_{s}$} & \colhead{$R_{200}$} & \colhead{c} & 
\colhead{$V_{200}$} & \colhead{$\rho_{0}$} & \colhead{$R_{c}$} & 
\colhead{$\chi^{2}_{pl}$\tablenotemark{d}} & \colhead{$\chi^{2}_{NFW}$\tablenotemark{d}} & 
\colhead{$\chi^{2}_{iso}$\tablenotemark{d}} & \colhead{$\chi^{2}_{mod\_iso}$\tablenotemark{d}}\\
\colhead{} & \colhead{[\kms]} & \colhead{} & \colhead{[kpc]} & 
\colhead{[kpc]} & \colhead{} & \colhead{[\kms]} & 
\colhead{[\gpercmcu]} & \colhead{[kpc]} & \colhead{} & \colhead{} & 
\colhead{} & \colhead{} } 

\startdata
NGC~2976\tablenotemark{a} & 52.1 & 0.27 & \nodata\tablenotemark{b} & \nodata\tablenotemark{b} & \nodata\tablenotemark{b} & \nodata\tablenotemark{b} & $1.3 \times 10^{-23}$ & 1.5 & 0.33 & $>5.8$ & 0.48 & 0.93 \\
NGC~4605\phm{$^{a}$} & 65.0 & 0.89 & $>42.0$       & $>347$ & $<8.3$  & $>246$ & $4.0 \times 10^{-23}$ & 0.7 & 8.28 & $<9.28$ & 1.53 & 4.92 \\
NGC~5949\phm{$^{a}$} & 63.3 & 0.93 & $>38.0$       & $>325$ & $<8.6$  & $>231$ & $2.2 \times 10^{-23}$ & 1.1 & 1.50 & $<2.00$ & 1.00 & 0.64 \\
NGC~5963\phm{$^{a}$} & 87.8 & 1.41 & $\phm{>}2.86$ & $\phm{>}102.7$ & $\phm{>}35.9$    & \phn72.9 & $1.3 \times 10^{-22}$ & 0.4 & 6.88 & $\phm{>}3.57$ & 1.22 & 3.04 \\
NGC~5963\tablenotemark{c} & 89.5 & 1.62 & $\phm{>}3.42$ & $\phm{>}109.9$ & $\phm{>}32.1$    & \phn78.0 & $1.2 \times 10^{-22}$ & 0.4 & 6.52 & $\phm{>}2.64$ & 1.07 & 1.82 \\
NGC~6689\phm{$^{a}$} & 53.1 & 1.07 & $\phm{>}41.5$ & $\phm{>}286.7$ & $\phm{>}\phn6.9$ & 203.6    & $1.4 \times 10^{-23}$ & 1.1 & 0.32 & $\phm{>}0.31$ & 0.95 & 0.42 \\

\enddata
\label{fitdata2}

\tablenotetext{a}{Using data from \citet{s03}.}
\tablenotetext{b}{No meaningful NFW fits could be made for NGC~2976,
even with only one parameter free.  The ``best'' fit had $\chi^2_{red} = 5.8$.}
\tablenotetext{c}{Using the rotation curve presented in this paper
  plus the rotation velocities measured by \citet*{bosma88} at
  $55\arcsec \le r \le 165\arcsec$.}
\tablenotetext{d}{All $\chi^{2}$ values listed here actually represent
the reduced $\chi^{2}$ ($\chi^{2}$ per degree of freedom in the fits).}

\end{deluxetable*}

\subsection{Alternative Density Profile Fits}
\label{otherfits}

In order to obtain a more complete understanding of the allowed shapes
of the inner density profiles of the galaxies in our sample, we also
fit the rotation curves with several other functional forms.  The two
most commonly used profiles in the literature are the
pseudo-isothermal profile (an isothermal sphere with a central core)
and the NFW profile\footnote{For our NFW fits, we chose the scale
radius $r_{s}$ and the virial radius $r_{200}$ to be the two free
parameters.}.  The functional form for the pseudo-isothermal density
profile is
\begin{equation}
\rho(r) = \frac{\rho_{0}}{1 + (r/r_{c})^{2}} ,
\label{iso}
\end{equation}

\noindent
where $\rho_{0}$ is the central density and $r_{c}$ is the core
radius \citep[e.g.,][]{kent}.

Since pseudo-isothermal density profiles are not expected on
theoretical grounds, and there is little observational support for NFW
profiles, it is important not to limit ourselves to only these
possibilities.  Fitting the data with other density profiles can help
to elucidate the physical meaning of the fits.  We therefore also used
a profile we defined to have characteristics intermediate between
those of the pseudo-isothermal and NFW profiles:
\begin{equation}
\rho(r) = \frac{\rho_{c}}{(r/r_{c})^{1/2}
(1 + r/r_{c})^{3/2}} \mbox{ }, 
\label{modiso}
\end{equation}

\noindent
where $r_{c}$ and $\rho_{c}$ are the characteristic radius and
density.  The inner slope of this profile is $\alpha = 0.5$, giving it
a shallow inner cusp.  At large radii the slope is $\alpha = 2$
(corresponding to a flat rotation curve), so the fact that our
observations only just reach the flat part of the rotation curve
should not adversely affect the fits.  Note that the $\chi^{2}$ values
for fits to this function are given in Tables \ref{fitdata1} -
\ref{fitdata2} even though we do not list $r_{c}$ and $\rho_{c}$ for
each galaxy.

We used a $\chi^{2}$-minimization routine to find the best fit
parameters for each of these functions.  With the exception of
NGC~4605, in which small-scale bumps and wiggles in the rotation curve
inflate $\chi^{2}$, typical reduced $\chi^{2}$ values for the best
fits were $\approx 0.3$, suggesting that the systematic uncertainties
on our rotation curves have been overestimated by a factor of about
1.7.  \citet{dutton} similarly inferred that our rotation curve error
bars for NGC~2976 were overestimated, indicating that the bootstrap
and Monte Carlo analysis described in \S \ref{ringfit} may be
excessively conservative.

The results of these fits after removing the contribution of the
stellar disk of each galaxy are summarized in Table \ref{fitdata1},
and fits without removing the stellar disk rotation velocities are
given in Table \ref{fitdata2}.  The best power law, NFW, and
pseudo-isothermal fits (after subtracting the stellar disk) are
displayed in Figures \ref{rcfig4605}b - \ref{rcfig6689}b by the green,
cyan, and magenta curves, respectively.  Residuals from each fit are
shown in the same colors in the lower right panels of each figure.

With the exception of NGC~5963 (and NGC~6689 in the minimum disk
case), we were unable to obtain reasonable NFW fits to the rotation
curves.  Since our data generally do not extend well into the flat
part of the rotation curve, $r_{s}$ and $r_{200}$ are highly
covariant.  To prevent both parameters from running away to
unphysically large values for NGC~4605, NGC~5949, and NGC~6689, we
held $r_{s}$ fixed and only fit for $r_{200}$.  An infinite number of
such fits are possible, with $\chi^{2}$ decreasing as $r_{s}$ and
$r_{200}$ increase.  The NFW fits listed in Tables \ref{fitdata1} and
\ref{fitdata2} are for $r_{s}$ values chosen to keep $\chi^{2}$ low
($\Delta \chi^{2} \approx 1$ relative to the best fitting power law)
without allowing $r_{200}$ to become completely unreasonable.

\subsubsection{Fit Results}

NGC~5963, strikingly, is best fit by a very steep ($\alphadm > 1$)
power law or an NFW profile unless its stellar mass-to-light ratio is
unrealistically high.  The formal best fit for NGC~4605 is a
pseudo-isothermal profile, although none of the profiles have good
$\chi^{2}$ values.  The best power law (which does not describe the
full rotation curve very accurately) still has a relatively steep
slope.  NGC~5949 is also best fit by pseudo-isothermal density
profile, but power laws with slopes slightly shallower than NFW, and
the modified pseudo-isothermal profile given in Equation \ref{modiso},
fit very well.  NGC~6689 is better described by a power law than a
pseudo-isothermal profile, and in the case of a low stellar M/L can
even be fit by an NFW profile, although the scale radius and virial
velocity suggested by the NFW fit are unrealistically large.  For
every galaxy except NGC~4605, Equation \ref{modiso} (the modified
pseudo-isothermal profile with a shallow central cusp) provides fits
that are comparable to or better than the pseudo-isothermal profile.

\subsubsection{\citet{n04} Density Profile}
\label{n04densityprof}

In addition to the profiles described above, we also fit the rotation
curves (after removing the stellar disks) with the new density profile
proposed by \citet{n04}:
\begin{equation}
\rho_{\eta}(r) = \rho_{-2}
e^{-\frac{2}{\eta}\left[(\frac{r}{r_{-2}})^{\eta} - 1\right]} ,
\label{nfw2}
\end{equation}

\noindent
where $\rho_{-2}$ and $r_{-2}$ are the density and radius at which the
logarithmic density profile slope equals $-2$ and $\eta$ is an
additional free parameter\footnote{\citet{n04} use $\alpha$ instead of
$\eta$, but since we have defined $\alpha$ to be the slope of a
power-law density profile we adopt a different symbol here for
clarity.} that controls how fast the density profile slope changes
with radius.  The rotation curve associated with this density profile
is
\begin{equation}
V(R) = \left[ \frac{4 \pi G\rho_{-2} e^{\frac{2}{\eta}}}{R\eta}
  \left(\frac{2}{\eta r_{-2}^{\eta}} \right)^{-\frac{3}{\eta}}
  \Gamma\left(\frac{3}{\eta}\right) \gamma\left( \frac{3}{\eta},
  \frac{2}{\eta}\left[\frac{R}{r_{-2}}\right]^{\eta} \right)
  \right]^{1/2} ,
\end{equation}

\noindent
where $\gamma(a,x)$ is the lower incomplete gamma function.  With the
extra free parameter afforded by this function, we are able to achieve
very good fits to the rotation curves of each of our five targets.  In
Table \ref{n04fitdata} we present the results of these fits and
compare them to the fits \citet{n04} performed to four simulated dwarf
galaxies that have total masses similar to our targets.  Only for
NGC~5963 are the fit parameters we derive remotely close to those
measured by \citet{n04} in their simulations.

\begin{deluxetable}{c c c c c}
\tablewidth{0pt}
\tablecolumns{5}
\tablecaption{Navarro et al. (2004) Profile Fit Results}
\tablehead{
\colhead{Galaxy} & \colhead{$r_{-2}$} & \colhead{$\rho_{-2}$} & 
\colhead{$\eta$} & \colhead{$\chi^{2}$\tablenotemark{a}} \\
\colhead{} & \colhead{[kpc]} & \colhead{[\gpercmcu]} & 
\colhead{} & \colhead{} } 

\startdata
NGC~2976\tablenotemark{b} & 1.65 & $5.6 \times 10^{-24}$ & 8.99\phn & 0.26 \\ 
NGC~4605\phm{$^{a}$}      & 1.34 & $5.8 \times 10^{-24}$ & 1.84\phn & 1.37 \\
NGC~5949\phm{$^{a}$}      & 3.85 & $7.5 \times 10^{-25}$ & 0.67\phn & 0.25 \\ 
NGC~5963\phm{$^{a}$}      & 6.44 & $6.0 \times 10^{-25}$ & 0.28\phn & 0.62\tablenotemark{c} \\
NGC~5963\tablenotemark{d} & 6.04 & $6.4 \times 10^{-25}$ & 0.28\phn & 0.43 \\
NGC~6689\phm{$^{a}$}      & 3325 & $1.3 \times 10^{-28}$ & 0.12\phn & 0.31 \\
\\[1mm]
\hline
\\[1mm]
D1\tablenotemark{e} & 4.55 & $1.1 \times 10^{-25}$ & 0.164 & \nodata \\ 
D2\tablenotemark{e} & 4.28 & $1.5 \times 10^{-25}$ & 0.211 & \nodata \\
D3\tablenotemark{e} & 3.62 & $1.5 \times 10^{-25}$ & 0.122 & \nodata \\
D4\tablenotemark{e} & 3.62 & $2.1 \times 10^{-25}$ & 0.166 & \nodata \\
\enddata
\label{n04fitdata}

\tablenotetext{a}{$\chi^{2}$ values listed here actually represent
the reduced $\chi^{2}$ ($\chi^{2}$ per degree of freedom in the fits)}.
\tablenotetext{b}{Using data from \citet{s03}.}
\tablenotetext{c}{The three-parameter fit using only our data for
  NGC~5963 is degenerate.  Better $\chi^{2}$ values than shown here
  can be achieved, but only when $r_{-2}$ and $\rho_{-2}$ run away to
  unrealistic values.  We therefore fixed $\eta$ to the value obtained
  when the \citet*{bosma88} data are included as well (see the
  following line of this table) and ran the fit with only two free
  parameters.}
\tablenotetext{d}{Using the rotation curve presented in this paper
  plus the rotation velocities measured by \citet*{bosma88} at
  $55\arcsec \le r \le 165\arcsec$.}
\tablenotetext{e}{Simulated dwarf galaxy halos from \citet{n04}.}

\end{deluxetable}

\subsection{Comparison to Previously Published Rotation Curve Data}
\label{oldrcs}
In the remainder of this section, we discuss details of our analysis
of each galaxy.  Readers who are primarily interested in the more
general results of our work may wish to skip to \S \ref{discussion}.

Each of the four galaxies we observed have existing \ha\ or \hi\
rotation curve data in the literature.  Our new data should improve on
previous measurements in velocity resolution, control of systematics,
angular resolution (in the case of \hi\ measurements), and
sensitivity, but it is useful as a first-order test of the accuracy of
our data to compare the rotation curves we obtain with previous
measurements.

NGC~4605 has been the subject of several rotation curve studies
\citep{rubin80,sofue98,us}.  All three of those studies derived very
similar rotation curve shapes and amplitudes.  While our methodology
is different from that employed by the previous authors (we do not
produce separate fits for the approaching and receding sides of the
galaxy), our results are qualitatively in agreement.  In particular,
the break in the rotation curve at a radius of 40\arcsec\ is quite
apparent in the two recent data sets (the \citealt*{rubin80} rotation
curve does not have enough angular resolution), and the maximum
rotation velocities match well.  The primary quantitative discrepancy
is the result of the incorrect plate scale assumed by \citet{us} for
the \ha\ spectrum they obtained.  In that paper, we used a value of
$3.0$~arcsec~pixel$^{-1}$, but subsequent measurements showed that the
actual plate scale is $2.1$~arcsec~pixel$^{-1}$.  Correcting this
error reduces the extent of the \ha\ rotation curve by 30\%, but
should not substantially change the power law indices measured for the
rotation curve and density profile.  Indeed, our new rotation curve
matches the old one near the center of the galaxy ($r <
10\arcsec$) and at large radii ($r > 65\arcsec$).  From 10\arcsec\ to
30\arcsec\ our rotation curve is a few \kms\ lower than the old one,
and from 30\arcsec\ to 65\arcsec\ our rotation curve exceeds the old
one by $1-6$ \kms.  The net effect of these changes\footnote{Note that
the stellar disk we use is not the same as that of \citet{us}.
\citeauthor{us} calculated the stellar rotation curve for an
exponential disk and we use the actual surface brightness profile to
determine the stellar rotation curve.  Although the resulting stellar
disks are not quite identical, the differences do not significantly
affect the density profile results.} is to modestly steepen the
maximum disk density profile from $\rhodm \propto r^{-0.65}$
\citep{us} to $\rhodm \propto r^{-0.88}$.  Given typical uncertainties
on $\alphadm$ of 0.1 in our analysis \citep{s03}, these results are
marginally consistent.  The earlier study by \citet{ps90} found a
density profile of $\rho \propto r^{-0.68}$, also consistent with
our results.  The best fitting power law rotation curve at large radii
still overestimates the rotation velocities in the inner region of the
galaxy, as seen in \citet{us} and Figure \ref{rcfig4605}a.

The rotation curve of NGC~5949 has previously been measured by
\citet{kp90} and \citet{courteau97} with long-slit spectra.  The
\citeauthor{kp90} data are consistent with the rotation curve we
derive, although our rotation velocities are somewhat larger and our
measurements extend to larger radii.  Our rotation curve matches that
of \citeauthor{courteau97} within the uncertainties of the data.

The kinematics of NGC~5963 have been investigated in detail in two
previous papers \citep{roman,bosma88}.  The shape of the \ha\ rotation
curve measured by \citet*{roman} is very similar to the shape of our
rotation curve; the amplitude they derive is slightly larger
(asymptotic rotation velocity of 131 \kms), but this appears to be the
result of the smaller inclination angle they used.  The \hi\ rotation
curve presented by \citet*{bosma88} is perfectly consistent with our
rotation curve over the common range of radii, except that the inner
two points of their rotation curve are affected by beam smearing.  For
the fits presented in Tables \ref{fitdata1} - \ref{n04fitdata}, we give
results for both our data alone, and our data with the \citet{bosma88}
points at large radii added.

The \ha\ velocity field of NGC~6689 was observed by the GHASP survey
\citep{garrido03}.  Those authors found a decidedly asymmetric
rotation curve, but the filter they used for the observations may have
cut off the emission line profiles on the approaching side of the
galaxy.  The receding side of the rotation curve presented by
\citet{garrido03} is consistent with the overall rotation curve we
derive out to a radius of 45\arcsec, but at larger radii our rotation
curve shows significantly smaller rotation velocities.

\subsection{Comments on Individual Galaxies}
\label{comments}

\subsubsection{NGC~4605}

\begin{figure*}[!ht]
\figurenum{8}
\epsscale{1.0}
\plotone{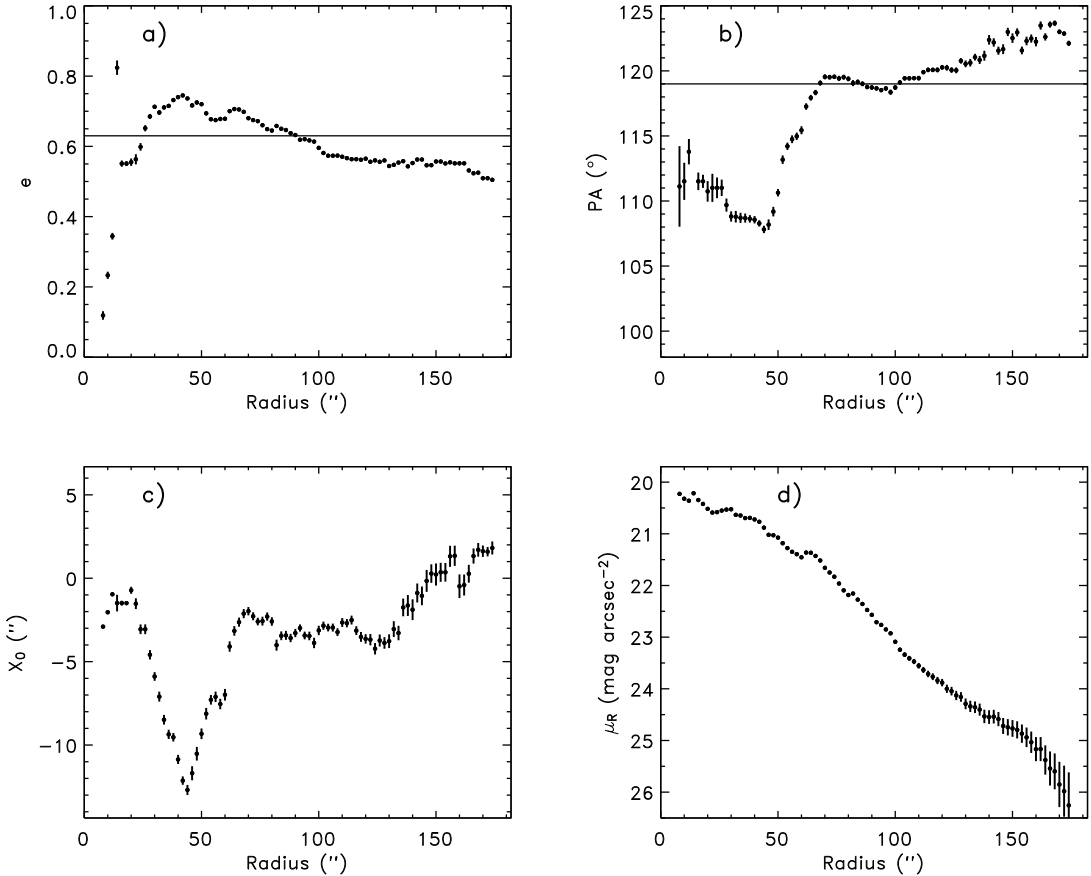}
\caption{$R$-band isophotal fit parameters for NGC~4605 with the
ellipticity, PA, and center left free to vary.  (a) Isophotal
ellipticity as a function of radius.  The peak at $r = 30 -
40$\arcsec\ is due to the asymmetric structure of the galaxy.  Note
the corresponding features in panels (b) and (c).  The horizontal line
shows the mean value used in our analysis.  (b) Isophotal position
angle as a function of radius.  The horizontal line shows the mean
value used in our analysis.  (c) Isophotal center position along the x
(right ascension) axis of the image, in arcseconds.  The position of
the nucleus corresponds to $X_{0} = 0$, which coincides with the
center of the outermost isophotes.  The center position along the y
(declination) axis shows almost identical behavior, but with smaller
amplitude because the asymmetric structure is located mostly east of
the galaxy center.  (d) Surface brightness as a function of radius,
using the parameters shown in the other panels.  Note that since our
final isophotal fits did not allow the parameters to vary with radius,
this surface brightness profile differs slightly from the one used to
derive the stellar disk rotation curve (Figure \ref{sbprofs}a).
\label{4605params}}
\end{figure*}

Of our four target galaxies, NGC~4605 is the only one that appears to
deviate significantly from axisymmetry.  The non-axisymmetric
structures are visible both photometrically and kinematically.  In
broadband optical images, the galaxy contains a pronounced elongated
region that is offset from its nucleus to the east.  This feature,
which has a somewhat higher ellipticity and a different position angle
than the rest of the galaxy, persists out to K-band although it
weakens with increasing wavelength.  In Figure \ref{4605params}, we
plot the isophotal fit parameters as a function of radius (where we
now allow the isophotal center, PA, and ellipticity to vary from ring
to ring), and the impact of the asymmetry is easily visible out to a
radius of 70\arcsec.  Since this lopsidedness is also apparent in our
\ha\ image (the galaxy's \ha\ emission extends to twice as large a
radius on its eastern side as on the western side) and CO map, it may
be associated with recent star formation.

The kinematic effects of this asymmetry are twofold: 1) the kinematic
center of the galaxy is offset from the nucleus, and 2) the galaxy
contains relatively large (and significant) noncircular motions.
Tilted-ring models of the galaxy centered on the nucleus result in a
systematic trend of the systemic velocity with radius.  Moving the
center position around so as to remove this trend and simultaneously
minimize the scatter of the systemic velocities about their mean
value, we found that the preferred kinematic center is southeast of
the nucleus by 10\arcsec.  Offsets of the kinematic center of this
magnitude do not significantly change the slope of the density
profile.  Irrespective of the center position, the models also show a
clear detection of radial motions in NGC~4605.  The amplitude of these
motions increases to 8 \kms\ at a radius of 20\arcsec, and then stays
constant out to 80\arcsec.  At larger radii (the outermost six rings)
the radial motions begin to increase again, reaching 20 \kms\ at the
edge of the galaxy ($r = 113\arcsec$).  However, this increase is not
very well constrained due to the poor angular coverage of the velocity
data at large radii.  Unsurprisingly, we find small ($2-6$ \kms) but
consistently nonzero $\cos{2\theta}$, $\sin{2\theta}$,
$\cos{3\theta}$, and $\sin{3\theta}$ terms in the velocity field of
NGC~4605 at various ranges of radii between 16\arcsec\ and 68\arcsec.

The combination of photometric and kinematic lopsidedness suggests
that the mass distribution of the galaxy indeed is lopsided, as
opposed to the possibility of asymmetric star formation mentioned
earlier.  Despite the lopsidedness of NGC~4605, the stellar disk is
described well by an exponential profile, as shown in Figure
\ref{sbprofs}a.

In order to determine the potential impact of these asymmetries on our
density profile analysis, we considered the maximum changes they could
cause.  The photometric structure makes it difficult to choose an
ellipticity and position angle that are representative of the whole
galaxy.  In \S \ref{stellardisks} we used a weighted mean of the
ellipticity and PA at radii between the outer edge of the asymmetric
structure (at $r = 70\arcsec$) and the edge of the galaxy.  As an
alternative, we tried setting the ellipticity to its maximum and
minimum values over the whole galaxy, and repeated the tilted-ring
modeling and isophotal fitting.  We found that the rotation curve and
density profile shape are largely insensitive to changes in the
inclination angle.  Even for ellipticities as small as 0.56 ($i =
66\degr$) and as large as 0.73 ($i = 79\degr$), the largest change in
the derived value of $\alphadm$ is $-0.05$.  We therefore conclude
that our analysis is robust to the uncertainties introduced by the
asymmetry of NGC~4605.

\subsubsection{NGC~5949}

The surface brightness profiles of NGC~5949 are very similar to those
of NGC~2976 \citep{s03}.  Both galaxies contain a nucleus, a shallow
(large scale length) inner disk, and a steep (small scale length)
outer disk.  Although other disk galaxies are known to have similar
structures \citep[e.g.,][]{pohlen02}, the origin and physical
significance of these two apparently distinct disks are not
understood.  Nevertheless, there is no reason to doubt that our
thin-disk approximation is valid for this galaxy, so the shape of the
stellar rotation curve we calculated should be accurate.  Due to the
high level of symmetry of NGC~5949 in all bands, the photometric
inclination and PA are extremely well constrained, leading to very
small systematic uncertainties on the rotation curve.  It is worth
noting that NGC~5949 is the only galaxy in our sample in which the
shape of the stellar rotation curve is very similar to the shape of
the observed rotation curve.  Therefore, it is possible to model this
galaxy with essentially only a stellar disk and barely any
contribution from dark matter.  The stellar mass-to-light ratio
required to accomplish this is $\mslr \approx 2.7 \mlr$, 30 \% higher
than the derived maximum disk value.

Our tilted-ring modeling of NGC~5949 reveals a small radial component
to the velocity field with modest significance.  These radial motions
are detected beginning at a radius of 20\arcsec, and remain present
out to the edge of the galaxy.  The maximum amplitude is 8 \kms, and
there are a total of six consecutive rings that have radial motions
deviating from zero by at least $1\sigma$.  Since the radial motions
are $\lesssim 10 \%$ of the rotation speed at all radii, they should
not affect our mass models of the galaxy.  Note that because we do not
know which side of the galaxy is closer to us, we cannot determine
whether the detected radial motions correspond to inflowing or
outflowing material.  We also detect a $\sim2$ \kms\ $\cos{2\theta}$
term from 24\arcsec\ to 36\arcsec\ and a $\sim3$ \kms\ $\sin{3\theta}$
term from 12\arcsec\ to 32\arcsec.

\subsubsection{NGC~5963}
\label{n5963}

Unlike the other three galaxies, the luminous component of NGC~5963
does not contain an easily identifiable exponential disk.  At the
center of the galaxy is a bright, elongated bar-like feature about
4\arcsec\ across.  Outside this source is a small disk-like region
(500 pc in radius) with four tightly wound spiral arms.  At a radius
of 15\arcsec\ (950 pc) the surface brightness profile begins a steep
decline, falling by nearly 3 magnitudes over 18\arcsec.  Surrounding
this region is an LSB, nearly exponential disk that extends out to a
radius of at least 120\arcsec\ (see Figure \ref{sbprofs}c).  The inner
spiral arms can be traced out into this outer disk at very low surface
brightness levels.

It is difficult to interpret this surface brightness profile in terms
of the standard model of a disk galaxy.  Ordinarily, one might assume
that the bright central region of NGC~5963 is a bulge that just
happens to be at the center of an unusually faint disk.  The clear
presence of spiral arms in this region, however, suggests that even
here the galaxy is highly flattened.  One possibility is that this
structure is a pseudobulge that has formed via secular evolution of
the galaxy \citep[e.g.,][]{kk04}.  The most sensible way to derive the
stellar rotation curve in this case is to apply our standard
assumption that the disk can be treated as infinitesimally thin.  The
assumption that the stellar mass-to-light ratio remains constant
throughout this galaxy may not be correct, but since the stars are far
from dominating the gravitational potential for reasonable
mass-to-light ratios this should not substantially change our results.

NGC~5963 also contains noncircular motions.  The radial term in the
tilted-ring fits deviates from zero by more than $1\sigma$ in five
rings, from 12\arcsec\ to 30\arcsec.  The radial motions reach as high
as 15 \kms\ in this region, before beginning to oscillate about zero
(with large uncertainties) at larger radii.  The second-order Fourier
terms in this galaxy both have amplitudes of $\sim3$ \kms\ from radii
of $16\arcsec-24\arcsec$ ($\sin{2\theta}$) and $12\arcsec-24\arcsec$
($\cos{2\theta}$).

\subsubsection{NGC~6689}

The high inclination of NGC~6689 makes our images and surface
brightness profiles less revealing of details of its structure.
Nevertheless, it is clear that this galaxy contains an exponential
disk and a central light excess that could be attributable to a
bulge or a bar.  No bar is evident in our images, but as we noted in
\S \ref{targets}, some catalogs classify the galaxy as barred.

Despite the large inclination angle, the observed \ha\ emission lines
from NGC~6689 are well represented by Gaussian fits.  This implies
that sightlines through NGC~6689 generally only intersect a single
\hii\ region, which is consistent with the appearance of the galaxy in
a narrow band \ha\ image.  Therefore, it is not necessary to use an
envelope-tracing method \citep[e.g.,][]{sofue,g04} instead of Gaussian
fits to extract rotation velocities from the spectra.

Neither our standard tilted-ring model nor the higher order harmonic
fits detected any deviations from circular rotation in NGC~6689.  To
some degree, this is likely due to the inclination of the galaxy,
which limits our resolving power along the minor axis where radial
motions are most prominent.  Nevertheless, a radial component of
$\sim10$ \kms\ as we found in the other three galaxies should have
been detected if present.

\section{DISCUSSION}
\label{discussion}

\subsection{Is There a Universal Density Profile?}
\label{universal}

The primary goal of this study is to determine as accurately as
possible the inner density profiles of the dark matter halos of the
galaxies we observed.  At radii of $\lesssim 1$ kpc, does the dark
matter density continue to increase, as predicted by NFW and numerous
other theoretical studies?  Or are the density profiles flat, with
constant density cores, as most previous observers have concluded?
Are the variations in density profile shape from galaxy to galaxy as
small as the simulations suggest?

\subsubsection{Observational Results}
For two of the galaxies in our sample (note that for the remainder of
the paper we include NGC~2976 in our analysis so that we have a total
sample of five galaxies), these questions are easy to answer.
NGC~2976, as \citet{s03} showed, unambiguously contains a constant
density core.  An $\alphadm = 0$ power law (constant density) provides
an excellent fit to the rotation curve.  A pseudo-isothermal profile
with a core comparable to the optical size of the galaxy also produces
a good fit.  NGC~5963, on the other hand, quite clearly has a very
steep central density profile.  A power law with a slope of $\alphadm
= 1.20$ fits the rotation curve very well, and an NFW fit with $r_{s}
\approx 11$ kpc and a concentration parameter of 14.9 is nearly as
good.  A pseudo-isothermal fit is significantly inferior.

For the remaining three galaxies, the interpretation of the fit
results is not nearly as straightforward.  NFW fits to the
disk-subtracted rotation curves can be carried out, but the fit
parameters are not usefully constrained.  Power laws with slopes
between $\alphadm = 0.78$ (NGC~4605) and $\alphadm = 0.88$ (NGC~5949)
fit the rotation curves well, but they are matched in each case by
pseudo-isothermal profile fits with comparable or better reduced
$\chi^{2}$ values.  

How can a rotation curve be simultaneously consistent with both a
pseudo-isothermal halo and an $\alpha \approx 0.8$ power law density
profile?  As shown by \citet{vdbs01}, the difference between the
rotation curves associated with these density profiles tends to be
smaller than typical observational uncertainties, even with the high
velocity resolution of our data.  The pseudo-isothermal profile has a
slope that varies continuously from 0 (constant density) at its center
to -2 (isothermal) at large radii.  A power law, of course, has a
constant slope that (for our fits) is intermediate between these two
values.  That both profiles fit the data suggests that 1) the mean
density profile slope over the observed region is equal to the value
preferred by the power law, but that 2) the slope changes with radius,
which is better described by the pseudo-isothermal profile.  Previous
studies have often assumed that if a pseudo-isothermal rotation curve
fits the data, then the galaxy in question must contain a
constant-density core.  Our successful fits with the modified
pseudo-isothermal profile (Equation \ref{modiso}) show that this
assumption is not correct.  Even when a pseudo-isothermal profile fits
well, other density profiles with steeper central slopes may provide
equally good fits.

Taken together, these observational results argue against the
proposition that all galaxies share a universal density profile.
Common two-parameter profiles other than a power law cannot fit all of
the galaxies in our sample, and even for power law fits the solutions
span the range from constant-density to very cuspy.  If a universal
density profile exists, the scatter from halo to halo is large.

\subsubsection{Simulation Results}

The idea of a universal dark matter density profile traces back to
NFW, and the most recent simulations continue to support this picture
\citep{n04,d04,stoehr}.  The favored functional form of the universal
profile, however, has not been agreed upon.  For several years the
debate focused on the exact value of the central cusp slope, but with
the increasing resolution of the simulations it now seems that this
question may be the wrong one to ask.  The highest resolution
$\Lambda$CDM simulations reveal that dark matter density profiles do
not converge to an asymptotic central slope \citep*{power03,n04,d04}.
Instead, the logarithmic slope continues to get shallower at smaller
radii, and the best procedure for extrapolating the profiles to radii
below the resolution limits\footnote{Note that the highest-resolution
simulations currently have a resolution limit of $500-1000$ pc, so
another factor of $\sim5$ improvement in the resolution is needed
before the simulations reach the scales probed by observations.} of
the simulations is not clear.

The new density profile proposed by \citet{n04} (Equation \ref{nfw2})
fits simulated dark matter halos more accurately and over a wider
radial range than does the original NFW profile.  All of the halos
presented by \citet{n04} are well described by the new profile.  A key
difference between this profile and the classic NFW or \citet{moore99}
profiles is that there is no well-defined central cusp.  The density
reaches a finite central value rather than diverging.  On
observationally relevant scales (100 to 1000 pc), though, the density
profiles remain rather steep.  Observations like ours with a
resolution of tens or hundreds of parsecs would be expected to find
density cusps only marginally shallower than an NFW profile.  Only on
\emph{sub-parsec} scales does the logarithmic slope of this profile
reach even $\alpha = 0.5$ \citep{n04}.

When the fits to Equation \ref{nfw2} for various halos are rescaled by
the characteristic density ($\rho_{-2}$) and radius ($r_{-2}$) of each
halo, they are all essentially identical, corroborating the hypothesis
of a universal density profile.  \citet*{d04} confirm the ability of
this functional form to fit their own independently simulated halos.
However, \citeauthor*{d04} also show that a generalized
three-parameter NFW profile with a formal central cusp is able to fit
the halos nearly as well.  Note that our observations still probe well
below the scales that are resolved by the simulations, and it is
unknown which, if any, of the profiles motivated by the simulations
provides the most accurate extrapolation to smaller radii.

\subsubsection{Comparison}

Despite our lack of knowledge about the precise functional form of
simulated density profiles at small radii, we can still compare our
observational results with the profiles that fit the simulations best.
In a general sense, it is obvious merely from inspection of Figures
\ref{rcfig4605} to \ref{rcfig6689} (and Figure 10 in \citealt{s03})
that the five rotation curves have rather different shapes.  This
visual impression is confirmed by the fits we performed in \S
\ref{results}.  If we use power laws to describe the density profiles,
the mean slope is $\alphadm = 0.73$, with a dispersion of 0.44.  By
comparison, \citet*{d04} find for the generalized NFW profile fits
that the central power law index is $\alphadm = 1.16 \pm 0.14$.  A
more dramatic (although less intuitive) illustration of the difference
between the observed and simulated density profiles is provided by our
fits with the new \citet{n04} density profile.  Although this formula
fits our data well, as shown in \S \ref{n04densityprof} and Table
\ref{n04fitdata}, we derive values of $\eta$ ranging from 0.12 to
8.99.  Even if we ignore the clearly absurd results for the flat
density profile of NGC~2976, we find a mean value of $\eta = 0.73$,
with a dispersion of 0.78.  From the simulations, \citet{n04} measure
$\eta = 0.172 \pm 0.032$ and \citet*{d04} are in agreement, finding
$\eta = 0.186 \pm 0.037$.  With the exception of NGC~5963, our fits
are seriously discrepant with these results.  If we force $\eta$ to
lie within the range preferred for the simulated halos, the scale
radius and scale density run away to unreasonable values (as can be
seen in Table \ref{n04fitdata} for NGC~6689).

Thus, there are two notable differences between our results and the
most recent CDM simulations: 1) the halo-to-halo scatter is $\gtrsim3$
times larger than the simulations, and 2) the observed central slopes
are on average shallower than the simulations.  Both of these
differences present challenges for future simulations.

Although the \emph{shapes} of the central density profiles we derive
disagree with the theoretical predictions, the actual \emph{values} of
the central densities are relatively consistent with the simulation
results.  In Table \ref{galaxyprops} we give measured values for the
dimensionless quantity $\Delta_{V/2}$ introduced by \citet{abw02} to
parameterize halo central densities.  In a $\Lambda$CDM cosmology, the
galaxies in our sample should have $\Delta_{V/2}$ values between
$10^{6}$ and $3 \times 10^{6}$ (assuming NFW density profiles),
similar to the values we observe.  The galaxies do have a tendency to
lie somewhat below the predicted level, but the difference is within
the $1\sigma$ scatter expected for $\Lambda$CDM.

\begin{deluxetable*}{c c c c c c c c c c c c}
\tablewidth{0pt}
\tablecolumns{12}
\tablecaption{Are Galaxy Parameters Correlated With Density Profile Slope?}
\tablehead{
\colhead{Galaxy} & \colhead{Distance} & \colhead{$M_{I}$} & \colhead{$\mu_{0,I}$\tablenotemark{a}} & 
\colhead{$V - I$} & \colhead{$M_{dyn}$\tablenotemark{b}} & \colhead{\mslr} & 
\colhead{$i$} & \colhead{$V_{max,halo}$\tablenotemark{c}} & \colhead{$R_{V/2}$\tablenotemark{d}} & 
\colhead{$\Delta_{V/2}$\tablenotemark{e}} & \colhead{$\alphadm$} \\
\colhead{} & \colhead{[Mpc]} & \colhead{[mag]} & \colhead{[mag arcsec$^{-2}$]} & \colhead{[mag]} & 
\colhead{[$M_{\odot}$]} & \colhead{[\mlr]} & 
\colhead{[deg]} & \colhead{[\kms]} & \colhead{[pc]} & \colhead{} &
\colhead{} }

\startdata
NGC~2976\tablenotemark{f} & 3.45 & $-18.6$ & 19.73 & 0.87    & $3.7 \times 10^{9}$  & 0.53  & 61.4 & 74 & 900 & $7.0\times10^{5}$ & 0.01  \\
NGC~4605 & 4.26 & $-18.9$ & 19.36 & 0.75    & $4.7 \times 10^{9}$  & 1.01  & 71.5 & 74 & 560 & $1.7\times10^{6}$ &  0.78  \\
NGC~5949 & 14.0 & $-19.8$ & 19.75 & 0.95    & $1.1 \times 10^{10}$ & 1.64  & 64.6 & 74 & 880 & $6.8\times10^{5}$ &  0.88  \\
NGC~5963 & 13.0 & $-19.1$ & 18.10 & 0.85    & $1.4 \times 10^{10}$ & 1.24  & 48.4 & 114 & 660 & $2.9\times10^{6}$ &  1.20  \\
NGC~6689 & 11.0 & $-19.0$ & 21.25 & \nodata & $1.5 \times 10^{10}$ & 1.96\tablenotemark{g}  & 76.0 & 94 & 1330 & $4.8\times10^{5}$ & 0.79  \\
\enddata
\label{galaxyprops}
\tablenotetext{a}{Extrapolated central surface brightness of the disk.}
\tablenotetext{b}{We calculate the dynamical mass as $M_{dyn} =
V_{max}^{2}R_{max}/G$, where $V_{max}$ is the maximum observed
rotation velocity and $R_{max}$ is the largest radius our observations
reach.  Since the galaxies extend to much larger radii, these numbers
clearly represent lower limits to the actual masses of each galaxy.}
\tablenotetext{c}{The maximum rotation velocity of the dark matter
halo after removing the stellar disk.}
\tablenotetext{d}{The radius at which the dark matter rotation curve
reaches half of $V_{max,halo}$, as defined by \citet*{abw02}.}
\tablenotetext{e}{The mean dark matter density within $R_{V/2}$,
in units of the critical density.}
\tablenotetext{f}{Using data from \citet{s03}.}
\tablenotetext{g}{This mass-to-light ratio is calculated for the Sloan
$r^{\prime}$ band, not the Kron-Cousins $R$ band.}

\end{deluxetable*}

\subsection{The Significance of NGC~5963}
\label{ngc5963}

Despite these differences from the simulations, the density profile of
NGC~5963 appears to be in good agreement with the theoretical models.
The rotation curve of this galaxy rises so rapidly than an NFW profile
fits the data very well.  A power law with a slope steeper than NFW
provides an even better fit.  To our knowledge, NGC~5963 is the only
low-mass disk galaxy for which a CDM-like central density cusp is
confirmed and shallow density profiles are ruled out.

Since NGC~5963 represents the exception rather than the rule, the
question is, are all galaxies formed with cuspy density profiles and
most lose them, or did an unusual event during the evolution of
NGC~5963 caused its density profile to become so steep?  The key to
answering this question is identifying what about NGC~5963 makes it
unique.

A number of galaxy properties from our sample are summarized in Table
\ref{galaxyprops}.  Among these galaxies, NGC~5963 has the highest
central surface brightness, despite its very low surface brightness
outer disk.  It also has the most unusual surface brightness profile
(see \S \ref{n5963}), although the significance of this is not clear.
NGC~5963 is at the brighter end of the luminosity range of our
targets, but the total span between the five galaxies is only $\sim1$
magnitude.  It does have the largest rotation velocity, and therefore
the highest mass, among our sample.  Other than mass, the most obvious
distinction between NGC~5963 and the other galaxies is the sharp
transition between the inner and outer surface brightness profiles.
This causes the derived stellar disk rotation curve to peak at small
radii ($r = 20\arcsec$) and then drop steeply, making the outer parts
of the galaxy highly dark matter dominated (see Figure
\ref{rcfig5963}b).  Although this photometric structure may be related
to the steep density profile of NGC~5963, we do not know its physical
origin.  If the galaxy really does contain a pseudobulge, then the
processes that have funneled baryons toward the center of the galaxy
could also have caused the dark matter to become more centrally
concentrated.

It is also noteworthy that high resolution two-dimensional velocity
fields and deep multicolor imaging are not necessary to recognize the
unique dark matter density profile of NGC~5963.  Using lower
resolution \hi\ data, a long-slit spectrum of modest resolution, and
photographic $U$ and $R$ band imaging more than 15 years ago,
\citet*{bosma88} already concluded that this galaxy probably contains
``an unusually centrally concentrated and massive halo''.  Since
high-resolution CDM simulations of galaxy structure had not yet been
done \citeauthor*{bosma88} could not generalize this result into any
broader implications for dark matter or galaxy formation.

\subsection{NFW and Pseudo-Isothermal Fits to Rotation Curves}

Our study is not alone in finding that pseudo-isothermal profiles fit
many rotation curves better than do NFW profiles.  A number of
previous authors have reached similar conclusions
\citep[e.g.,][]{db01a,bs01,db01b,dbb02,swb03,s03}.  We emphasize,
however, that the observation that pseudo-isothermal fits produce
lower reduced $\chi^{2}$ values than NFW or Moore profiles does not
demonstrate the presence of a constant-density core.  We showed in \S
\ref{otherfits} that in many cases power law density profiles with
intermediate ($0 < \alpha < 1$) slopes provide comparable fits.  A
modified pseudo-isothermal profile with an $\alpha=0.5$ central cusp
(Equation \ref{modiso}) also fits these rotation curves as well as a
true pseudo-isothermal profile with a constant-density core does.
Since current theories of galaxy formation do not lead naturally to
either cores or pseudo-isothermal density profiles, a preference for
constant-density cores over shallow density cusps is not justified.
Whether these galaxies actually contain a central region of constant
density or if they have shallow cusps cannot be determined from the
present data despite the high resolution of our measurements.

We do agree with past studies that for most galaxies the NFW form does
not fit the inner density profile very effectively (see Figures
\ref{rcfig4605}b-\ref{rcfig6689}b).  When $r_{s}$ and $r_{200}$ are
constrained to stay roughly in the range expected from simulations,
the $\chi^{2}$ values for NFW fits are generally rather high.  We also
point out, however, that the NFW fit parameters are quite poorly
constrained unless the observations extend well into the flat part of
the rotation curve.  If there are only a few data points on the flat
part of the rotation curve (especially when these data points have the
largest uncertainties, as is often the case), they do not provide
significant leverage on the fit.  In such cases, the NFW scale radius
and virial radius (or any other pair of parameters that can be used to
describe an NFW profile) become completely degenerate.  In order to
obtain accurate estimates of the NFW fit parameters, our results
suggest that $\gtrsim 50$ \% of the observed part of a galaxy must
have a nearly flat rotation curve.

\subsection{Implications for CDM}
\label{implications}

Combining this study with our previous analysis of NGC~2976, we have
shown that the five galaxies in our sample have density profiles with
a wide range of central slopes.  If we fit the data with power laws,
we find that three galaxies have central cusps of $\alphadm \approx
0.8$, while the other two galaxies have very different central slopes
($\alphadm = 0.01$ and $\alphadm = 1.20$).  Of the outliers, NGC~2976
is a satellite of M~81, and thus may have been subject to tidal
stripping.  There are suggestions in the literature that tidal
stripping acts to remove density cusps \citep{stoehr02,hayashi03}, but
the most recent simulations find that the central density slope is not
altered by tidal effects \citep{kaz04b}.  It is not clear why NGC~5963
differs so strongly from the other galaxies.  These fits demonstrate
that while galaxies with steep central density cusps do exist, they
are not shared by all galaxies.  Furthermore, no single value of the
central slope can describe all five of the rotation curves in our
sample.

On the other hand, if we consider the pseudo-isothermal fit results
(as most previous observational studies have done), we find that good
fits can be obtained for four of the five galaxies (see Table
\ref{fitdata1}).  It is interesting to note that these galaxies all
have central densities within a factor of seven of each other.  Only
for NGC~5963 is the quality of the pseudo-isothermal fit sufficiently
poor that a constant-density core can be excluded.  The rotation
curves of the other four galaxies are consistent with cores, but all
except NGC~2976 are also consistent with cusps.  The implication of
these findings is that some previous studies may have overestimated
the disagreement between their data and the CDM simulations.

Based on these results, we reach the following conclusions:  
\begin{itemize}
\item First, the observed variation in density profile slopes from
halo to halo is much larger than expected from the simulations.  We
also find that none of the commonly used density profile functional
forms can describe all five galaxies.
\item Second, most galaxies have density profiles significantly
shallower than the $\alphadm = 1$ central cusps preferred by the
simulations.  Nevertheless, the difference between the central slopes
we measure and the theoretical values is much smaller than suggested
by most previous studies.
\item And third, some galaxies do contain $\alphadm \gtrsim 1$ cusps
with an overall density profile that follows the NFW form, hinting
that it may still be possible to reconcile the results of the pure
dark matter cosmological simulations with observations.
\end{itemize}

Do these differences between the observations and the simulations
indicate a fundamental problem with the CDM paradigm?  Probably not.
A great many plausible theoretical explanations have been proposed in
attempts to understand this problem, and a few of the recent ideas are
mentioned below.  \citet{tn01} suggested based on phase-space
arguments that CDM density profiles should have $\alpha = 0.75$ cusps,
very similar to the average we measure.  \citet{ricotti} found that
the dark matter halos of simulated dwarf galaxies have shallower cusps
than those of massive galaxies.  In a similar experiment, however,
\citet{colin} reached the opposite conclusion, so the degree to which
dwarf galaxy density profiles are expected to match those of large
galaxies has not yet been settled.  \citet{mrbk} showed that major
mergers leave cuspy remnants unless both precursors had cores, so at
least in the case of pure dark matter halos, mergers cannot destroy
cusps.  Baryonic processes, however, probably can
\citep[e.g.,][]{wk02}.  Even the dark matter itself may be able to
flatten cuspy density profiles; \citet{cpma} argued that energy
deposition by merging dark matter substructures can puff up central
cusps into shallower density profiles.  This effect might have escaped
notice in previous simulations due to insufficient resolution or
overmerging (unrealistic destruction of subhalos in dense regions).
Finally, \citet{hayashi04} have shown that if dark matter halos are
significantly triaxial (see \S \ref{triaxialeffects}), for some viewing
angles the derived density profiles can appear much shallower than the
actual density profiles.

Since we have shown that cuspy density profiles are consistent with a
large fraction of our sample, and in light of the variety of potential
effects not currently accounted for in the simulations that could
change the predicted density profiles, there is little reason to
conclude at this point that the density profile controversy represents
a crisis for CDM.

\section{THE EFFECTS OF HALO TRIAXIALITY}
\label{triaxialeffects}

Almost every previous observational study of density profiles and
rotation curves has assumed that dark matter halos are spherical.  CDM
simulations, however, suggest that dark matter halos are triaxial
\citep{dc91,w92,cl96,js02}.  Unfortunately, very few observations of
individual galaxies are available to test this prediction.  Several
polar-ring galaxies have been studied, with results ranging from axis
ratios as small as 0.3-0.4 up to axis ratios of 1 (spherical halos)
\citep{swr83,wms87,ss90,sackett94,ca96,iodice}.  A variety of
techniques suggest that the gravitational potential of the dark matter
halo of the Milky Way is close to spherical \citep{kt94,om00}.
Despite the scarcity of reliable constraints on individual halo
ellipticities, statistical arguments provide a way to determine the
average shape of galaxy halos.  \citet{fdz92} show that the observed
scatter in the Tully-Fisher (TF) relation places a strong upper limit
on the allowed ellipticities of spiral galaxy disks.  Even if
\emph{all} of the TF scatter is caused by elongated disks, the mean
disk ellipticity is required to be less than 0.1.  More likely, the
scatter is a result of a combination of effects, indicating that on
average disk ellipticities are 0.05 or smaller.  In this section we
consider the effects triaxial halos could have on the velocity fields
we observe.

\subsection{Halo Oblateness}

Galaxy disks are expected to be oriented such that the disk lies in
the plane described by the major and intermediate axes of the halo,
with the $z$-direction pointing along the halo minor axis
\citep{sharma,bailin}.  We first consider the effect of this
flattening of the halo, under the assumption that halo shape in the
disk plane is circular.

The rotation velocity of a disk in such an oblate spheroidal halo is
given by Equation 2-91 in \citet{bt87}:
\begin{equation}
v_{rot}^{2}(R) = 4\pi G \sqrt{1 - e^{2}} \int_{0}^{R} 
  \frac{\rho(m) m^{2} dm}{\sqrt{R^{2} - m^{2} e^{2}}} ,
\label{oblateeqn}
\end{equation}

\noindent
where $\rho(m)$ is the density profile of the halo in the spheroidal
coordinate $m$ (defined by $m^{2} = r^{2} + z^{2}/(1-e^{2})$) and $e$
is the eccentricity of the spheroid.  For any density profile, this
integral can be evaluated numerically to give the rotation curve for
various degrees of oblateness.  We performed this calculation for
spheroidal power laws ($\rho (m) = \rho_{0} (m/m_{0})^{-\alpha}$) and
NFW profiles ($\rho (m) = \rho_{c}\delta_{c}(m/m_{s})^{-1}(1 +
(m/m_{s})^{2})^{-1}$).  The change from a spherical halo to an oblate
one could affect both the shape of the rotation curve and its
amplitude.  We find that as the halo becomes flatter ($e \rightarrow
1$) the amplitude of the rotation curve increases (as expected, since
the mass becomes concentrated closer to the disk).  We also find that
the shape of the rotation curve is completely independent of the halo
flattening for a power law; the same power law index for the rotation
curve (and density profile) is derived for any value of $e$.  For an
NFW density profile the shape of the rotation curve changes very
subtly with eccentricity (the peak of the rotation curve shifts to
smaller radii as the halo becomes flatter), but the inner slope of the
rotation curve is essentially unaffected.

\subsection{Disk Ellipticity}
\label{diskellip}

In the previous subsection, we showed that the flattening of the halo
does not alter the observed density profile for a circular disk.  We
now remove the assumption that the disk is circular and study the
effects of disk ellipticity on the observed velocity field.

\subsubsection{Are Noncircular Motions Common in Disk Galaxies?}
\label{noncirc}

Four of the five galaxies in our sample show evidence for a radial
component to their velocity fields.  In two of these, the radial
motions are detected very strongly; for the other two the significance
of the radial term is lower, although it is still confirmed at
$\sim95$ \% confidence.  Other recent studies have also begun to find
significant numbers of galaxies with noncircular motions
\citep*{schoen97,swaters03,coccato,blo04,wbb04}.  A large majority of
the galaxies studied by these authors (and us) are not barred.
However, of the galaxies in which observations could have detected
radial motions, nearly all indeed appear to contain them.  Are radial
components to the velocity fields ubiquitous in late-type spiral
galaxies?  What is the cause of these motions?

\subsubsection{Measuring Disk Ellipticity With Noncircular Motions}

Since there is no strong evidence for bars in the four galaxies where
we detect noncircular motions, other sources of these motions should
be considered.  First, we note that the observed noncircular motions
are dominated by the radial component.  The simplest possibility is
that the galaxies contain strong radial flows directed towards their
centers.  Given the observed magnitude of the radial motions at a
radius of $\sim1$ kpc and an assumed volume density for the gas (1 H
atom cm$^{-3}$), if the radial motions actually represent a net inflow
of gas we conservatively estimate that all of the gas would accumulate
within a $\sim1$ kpc radius of the galaxy centers in 1-3 Gyr.  The
star formation rate over the same region is at least an order of
magnitude too small to consume the inflowing gas.  We therefore
conclude that this interpretation is not viable.  The most intriguing
remaining explanation is that the gas is moving on elliptical orbits,
which could result from the influence of a triaxial dark matter halo.

In a disk galaxy with a triaxial dark matter halo, the potential in
the plane of the disk will in general be elliptical.  The closed
orbits in such a potential are also ellipses, which means that the
observed line-of-sight velocities of an elliptical disk will differ
from pure circular rotation.  This problem has been considered in
detail by \citet{binney}, \citet{teuben91}, \citet{fvgdz},
\citet*{schoen97}, and \citet{schoen98}.  \citet*{fvgdz} showed that an
elliptical potential induces components in the velocity field
proportional to the ellipticity and the angle between the long axis of
the ellipse and the observer's line of sight.  Adopting their
formalism (\citealt*{fvgdz}, Equation A10), we have the following
relations:
\begin{eqnarray}
\label{c1}
\hat c_{1} = \left[ 1 - (\frac{3}{4} - a) \epsilon_{R} \cos{2\phi_{obs}} \right]v_{c}\sin{i} \\
\hat s_{1} = \left[\left( \frac{5-q^{2}}{4(1-q^{2})} - a\right) \epsilon_{R} \sin{2\phi_{obs}}\right]v_{c}\sin{i} ,
\label{s1}
\end{eqnarray}

\noindent
where $\hat c_{1}$ and $\hat s_{1}$ are the $\cos{\theta}$ (rotation)
and $\sin{\theta}$ (radial) components of the tilted-ring model,
respectively.  The hat symbols indicate that these quantities are
derived under the assumption of circular orbits, and will therefore
differ systematically from the true values if the potential is
elliptical.  These formulae are valid when the rotation curve,
$v_{_c}(r)$, can be represented as a power-law with index $\beta$,
which we have shown is a reasonable assumption for these galaxies.
$\beta$ is related to $a$ by $a = \case{1}{2}\beta/(1 + \beta)$.  $q$
is the axis ratio of the galaxy, which we take from the photometry,
$\phi_{obs}$ is the angle between the long axis of the potential and
the line of sight, and $\epsilon_{R}$ is the ellipticity of the orbit.
The relationship between the ellipticity of the potential,
$\epsilon_{pot}$, and that of the orbit, $\epsilon_{R}$, is given by
\citet*{fvgdz}:
\begin{equation}
\epsilon_{pot} = \frac{1 - \beta}{1 + \beta} \epsilon_{R} .
\label{potentialeqn}
\end{equation}

In order for us to make the simplifying assumption that the apparent
rotation velocities are equal to the true rotation velocities (i.e.,
that the orbits are nearly circular), the following condition must be
met:
\begin{equation}
(\case{3}{4} - a) \epsilon_{R} \cos{2\phi_{obs}} \ll 1.
\label{circcondition}
\end{equation}

\noindent
In this case, Equation \ref{c1} reduces to $\hat c_{1} =
v_{c}\sin{i}$.  Substituting this expression into Equation \ref{s1}
and solving for the potential ellipticity yields
\begin{equation}
\epsilon_{R} \sin{2\phi_{obs}} =
\frac{1}{\left(\frac{5-q^{2}}{4(1-q^{2})} - a\right)} 
\frac{\hat s_{1}}{\hat c_{1}}.
\label{soln}
\end{equation}

\begin{figure*}[!ht]
\figurenum{9}
\epsscale{1.0}
\plotone{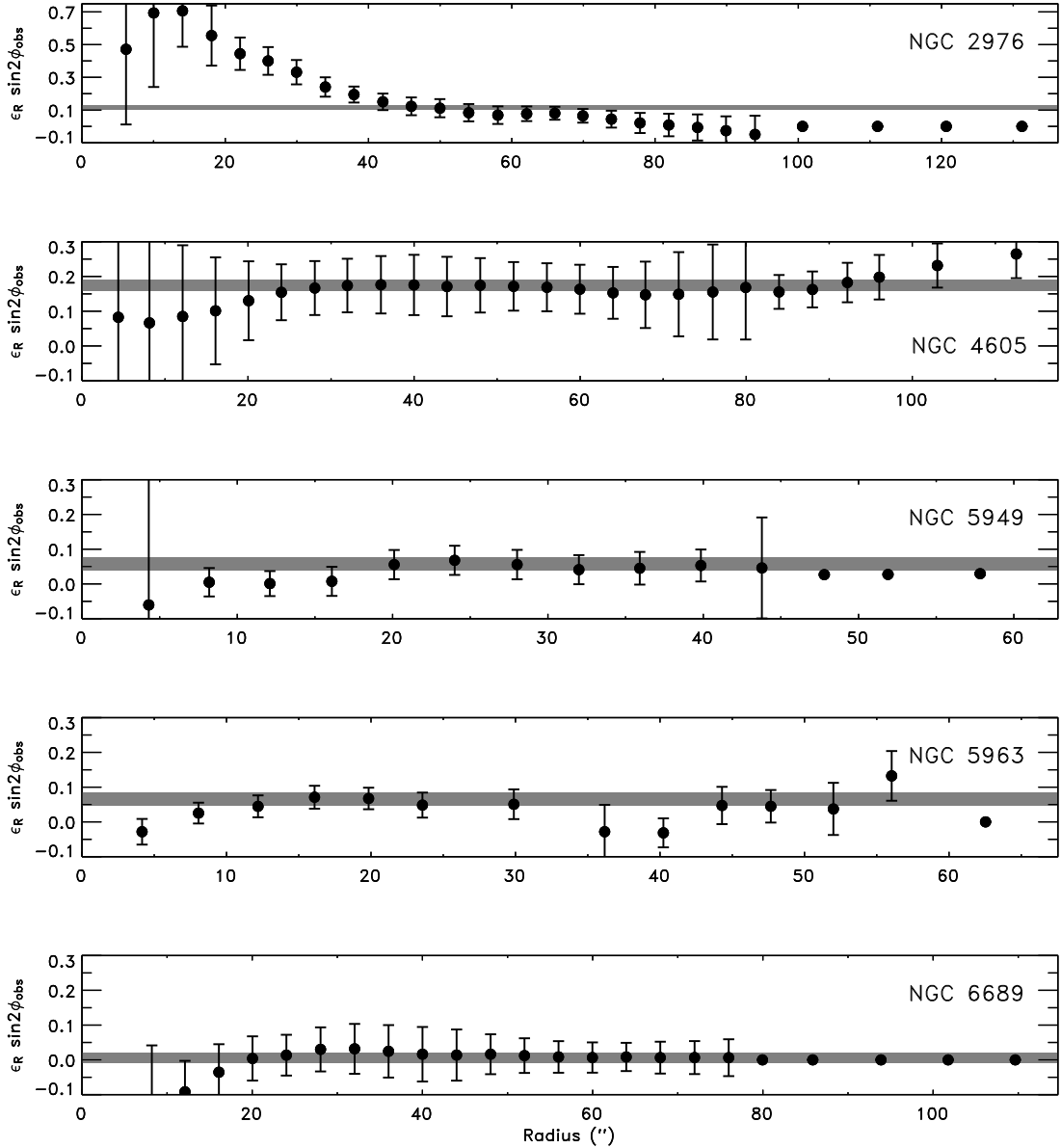}
\caption{$\epsilon_{R} \sin{2\phi_{obs}}$ as a function of radius for
(from top to bottom) NGC~2976, NGC~4605, NGC~5949, NGC~5963, and
NGC~6689.  The shaded gray bands across each panel represent the
weighted mean value of $\epsilon_{R} \sin{2\phi_{obs}}$.  The points
at large radii without error bars correspond to rings in which the
radial velocity was held fixed because the ring contained too few data
points or the angular coverage of the data was not sufficient to
constrain the radial motions.  These points do not contribute to the
weighted average and should not be considered to contain significant
information about the orbital ellipticity.  Only in NGC~6689 does the
gas appear to be traveling on circular orbits, although it is also
possible that $\phi_{obs} \approx 0\degr \mbox{ or } 90\degr$ for this
galaxy.  The ellipticity of the potential is a factor of a few smaller
than the ellipticity of the orbits that is plotted here (see Equation
\ref{potentialeqn}).  Note that these values are lower limits on the
true ellipticity due to the unknown viewing angle.
\label{ellipticity}}
\end{figure*}

\noindent
Thus, if the observed noncircular motions are indeed due to an
elliptical potential in the plane of the disk, the amplitudes of the
circular and radial motions put a direct constraint on the
ellipticity.  Note that since $\sin{2\phi_{obs}} \le 1$ this
constraint is actually a \emph{lower limit} on $\epsilon_{R}$.  Using
the known values of $q$ and $a$, and the results of our tilted-ring
models for $\hat c_{1}$ and $\hat s_{1}$, we calculate $\epsilon_{R}
\sin{2\phi_{obs}}$ as a function of radius for each galaxy.  The
results are shown in Figure \ref{ellipticity}.  NGC~4605, with its
very strongly detected radial motions, has a large and nearly constant
ellipticity that deviates significantly from zero beginning at a
radius of 20\arcsec.  The mean value of $\epsilon_{R}
\sin{2\phi_{obs}}$ over the entire galaxy is $0.175 \pm 0.016$, making
the disk of this galaxy substantially elliptical.  From Equation
\ref{potentialeqn}, we see that $\epsilon_{pot} = 0.18\epsilon_{R}$,
so the ellipticity of the potential is at least 0.03.  $\epsilon_{R}
\sin{2\phi_{obs}}$ is detected at the $\sim3\sigma$ level in NGC~5949
and NGC~5963.  We find that $\epsilon_{R} \sin{2\phi_{obs}} = 0.043
\pm 0.014$ for NGC~5949 and $\epsilon_{R} \sin{2\phi_{obs}} = 0.060
\pm 0.020$ for NGC~5963.  We do not detect any evidence for elliptical
orbits in NGC~6689: $\epsilon_{R} \sin{2\phi_{obs}} = 0.007 \pm
0.014$.  NGC~2976 is the only galaxy in which the ellipticity is not
constant with radius.  This could suggest that either the halo
structure is more complicated in this galaxy (e.g., the axis ratios of
the halo change with radius), or that its radial motions may be caused
by something other than triaxiality.  If the halo of NGC~2976 is
triaxial, the mean ellipticity of the disk orbits is $\epsilon_{R}
\sin{2\phi_{obs}} = 0.116 \pm 0.013$.

Given the determination of $\epsilon_{R} \sin{2\phi_{obs}}$, we can
now go back and confirm that the assumption stated in Equation
\ref{circcondition} (that the difference between the apparent and true
rotation velocities is small) was justified.  As long as the
ellipticity is modest ($\epsilon_{R} \lesssim 0.2$, which requires
that the viewing angle be larger than $\sim9\degr$ for most of these
galaxies), this condition will be satisfied.  Note, however, that for
NGC~4605 $\epsilon_{R}$ may be larger than this value.

These calculations show that, if the observed radial motions are
indeed due to halo triaxiality, the lower limit on the mean potential
ellipticity is of order a few percent.  Since the scatter in the TF
relation indicates that the average ellipticity of the potential has a
strict upper limit of 0.1 \citep{fdz92}, the ellipticities are now
constrained within a relatively narrow range.  CDM simulations predict
that the dark matter halos of galaxies similar in mass to our targets
should have axis ratios of $c/a = 0.47$ and $b/a = 0.62$ \citep{js02},
which would suggest larger ellipticities than we measure.
\citet{bailin} find somewhat less triaxial halos with $c/a = 0.65$ and
$b/a = 0.83$ for halos with masses of $\sim3 \times 10^{10}
M_{\odot}$.  Adding gas cooling to the simulations, however, may make
the inner regions of the halo much more spherical \citep{kaz04}.  With
observations of a larger sample of galaxies, it may be possible to
obtain robust constraints on the three-dimensional shapes of galaxy
halos, which would then provide another strong test of CDM simulations
on small scales.

\section{SUMMARY AND CONCLUSIONS}
\label{conclusions}

We have used two-dimensional CO and \ha\ velocity fields, sampled at
high spatial resolution and high spectral resolution, to constrain the
dark matter density profiles of four nearby, low-mass, late-type
galaxies.  We obtained rotation curves from the data cubes by
constructing tilted-ring models, and found that three of the four
galaxies contain radial motions in addition to rotation.  Combining
these data with the observations of NGC~2976 presented by \citet{s03}
gave us a sample of five galaxies to study.

For each galaxy we constructed a model of the stellar disk and
investigated the density profile of the dark matter halo under varying
assumptions about the stellar mass-to-light ratio.  In most cases,
over the range of plausible mass-to-light ratios, the dark matter
density profiles do not change significantly.  For the mass-to-light
ratios indicated by the galaxy colors, we found that the five galaxies
span a range of central density profile slopes from $\alphadm = 0$ to
$\alphadm = 1.20$.  The mean slope is $\alphadm = 0.73$, with a
dispersion of 0.44.  Neither of the standard density profile
functional forms (pseudo-isothermal and NFW) is able to adequately fit
all five galaxies.  The scatter in slope that we observe is three
times larger than that seen in the simulations, and the mean slope is
smaller than predicted.  We do note, however, that NGC~5963 is the
first low-mass disk galaxy in which a cuspy density profile of the
predicted form is confirmed, while shallower alternative profile
shapes are ruled out.

Although these results still indicate significant disagreements
between CDM simulations and the properties of real galaxies, there are
also some differences between the findings of this paper and those of
previous observational analyses.  Many authors have concluded that
density profiles with constant-density cores fit the rotation curves
of dwarf and LSB galaxies better than cuspy density profiles do
\citep*[e.g.,][]{db01a,bs01,db01b,dbb02,swb03,s03}.  Typical results
from these studies are that the distribution of density profile slopes
is $\alpha = 0.2 \pm 0.2$ \citep*{db01a,dbb02,dbbm03}.  Some other
studies agreed that cored profiles generally fit the data better, but
argued that this effect is primarily the result of systematic
uncertainties in the observations \citep*{vdb00,vdbs01,smvb03}.  When
we fit power laws to the rotation curves we find significantly steeper
density profile slopes on average than previous authors did, although
both pseudo-isothermal density profiles (constant-density cores) and
relatively cuspy profiles are consistent with our data.  Despite our
disagreement on the value of the mean slope, we agree with other
recent studies that found that dark matter density profiles span a
wide range of slopes \citep*[e.g.,][]{dbbm03,smvb03}.

What is responsible for the steeper slope that we find?  One important
element of our study is that the two-dimensional velocity fields, high
spatial resolution, and target selection we employed significantly
reduce the impact of systematic uncertainties on our results
\citep{s03}.  As an illustration of this point, \citet{vdbs01} showed
that to obtain tight constraints on density profile slopes, at least
one of the three following conditions must be met: 1) the
uncertainties on the rotation curve must be $\lesssim 0.2$ \kms, 2)
the rotation curve must extend out beyond 20\% of the virial radius,
and 3) the rotation curve must extend inward to radii less than
$\sim2$\% of the virial radius.  While we do not reach the velocity
precision of criterion (1), we approach it as closely as feasible,
with typical systematic uncertainties as small as $\sim2$ \kms.
Although higher velocity resolution spectroscopy is possible, our
velocity measurements have already run into the limit imposed by the
small-scale random motions of $\sim5$ \kms\ that most galaxies appear
to contain.  Without \hi\ observations at large radii, we also fail to
meet criterion (2), but our high spatial resolution does allow us to
satisfy criterion (3).  Galaxies with masses similar to those studied
here may have virial radii of 50 - 75 kpc \citep{bullock01,n04}, so
even for our most distant targets our $\sim250$ pc resolution probes
well within 2\% of the virial radius.  In addition, we explored
functional forms for the density profiles other than the standard
NFW/pseudo-isothermal dichotomy, thereby demonstrating that good
pseudo-isothermal fits in general do not rule out cuspy density
profiles.  Given these new results, we conclude that while the
discrepancies between CDM and galaxy rotation curves may not be as
severe as previously thought, improved simulations are needed in order
to determine whether the theoretical and observational findings can be
brought into agreement.

Finally, we considered the origin of the noncircular motions that are
now being detected regularly in disk galaxies.  A number of previous
authors have showed that a triaxial dark matter halo can produce
exactly the observed effect on the gas orbits.  Under the assumption
that halo triaxiality is causing the noncircular motions, we measure
lower limits to the orbital ellipticities of 0.02-0.17.  These
correspond to lower limits on the ellipticities in the potential
between 0.005 and 0.03.

\acknowledgements{This research was partially supported by NSF grant
AST-0228963.  We thank the anonymous referee for suggestions that
clarified the paper.  We also thank Kyle Dawson for obtaining images
of NGC~6689, and we appreciate the work of our WIYN telescope
operators Gene McDougall, Hillary Mathis, Andrew Alday, and Doug
Williams.  JDS gratefully acknowledges the assistance of Tim Robishaw
in making color figures in IDL, among other things.  We also thank
Chung-Pei Ma for providing comments on a draft of this manuscript and
acknowledge useful discussions with Kambiz Fathi.  This publication
makes use of data products from the Two Micron All Sky Survey, which
is a joint project of the University of Massachusetts and the Infrared
Processing and Analysis Center/California Institute of Technology,
funded by the National Aeronautics and Space Administration and the
National Science Foundation.  This research has also made use of the
NASA/IPAC Extragalactic Database (NED) which is operated by the Jet
Propulsion Laboratory, California Institute of Technology, under
contract with the National Aeronautics and Space Administration,
NASA's Astrophysics Data System Bibliographic Services, the SIMBAD
database, operated at CDS, Strasbourg, France, and the LEDA database
(http://leda.univ-lyon1.fr).}

\end{document}